\documentclass[preprint,review,12pt,3p]{elsarticle}
\usepackage{lineno,hyperref}
\modulolinenumbers[5]

\usepackage{sectsty}

\usepackage{amsmath,bm,enumitem}									% Math
\usepackage{float}
\usepackage{graphicx,epstopdf}
\usepackage{verbatim}
\usepackage{subfigure}
\usepackage{color}
\usepackage{multirow}
\usepackage{caption}
\begin{document}

\begin{frontmatter}

\title{Continuum model for dislocation structures of semicoherent interfaces}

\author{Luchan Zhang, Xiaoxue Qin}
\author{Yang Xiang\corref{mycorrespondingauthor}}
\address{Department of Mathematics, Hong Kong University of Science and Technology, Clear Water Bay, Kowloon, Hong Kong}

%\author[mymainaddress]{}

\cortext[mycorrespondingauthor]{Corresponding author}
\ead{maxiang@ust.hk}

%\address[mymainaddress]{Department of Mathematics, Hong Kong University of Science and Technology, Clear Water Bay, Kowloon, Hong Kong}

\begin{abstract}
In order to relieve the misfitting elastic
energy, the hetero-interfaces become semicoherent by forming networks of dislocations. These microscopic
structures strongly influence the materials properties associated with the development of advanced materials.   We  develop a continuum model for the dislocation structures of semicoherent interfaces.
The classical Frank-Bilby equation that governs the dislocation structures on
semicoherent interfaces is not able to determine a unique solution. The available
methods in the literature either use further information from atomistic simulations or
consider only special cases (dislocations with no more than two Burgers vectors) where the Frank-Bilby equation has a unique solution.
In our continuum model,
the dislocation structure of a semicoherent interface is obtained by minimizing the
energy of the equilibrium dislocation network with respect to  all the possible Burgers vectors, subject to the constraint of the Frank-Bilby equation. The continuum model is validated by comparisons with atomistic simulation results.

%Comparisons with atomistic simulation results show that
%our continuum model is able to give excellent predictions of dislocation structures on semicoherent interfaces.
\end{abstract}


\end{frontmatter}

%\linenumbers

\section{Introduction}

The interfaces between different materials or different phases commonly form semicoherent structures that consist of discrete dislocation networks to accommodate the lattice misfit between the two materials~\cite{Merwe1950,Matthews1966,Sutton1995}. Such semicoherent interfaces
play essential roles in the mechanical, electronic and plasticity properties that are associated with the development of novel composite materials and alloys \cite{Merwe1950,Matthews1966,Sutton1995,Quek2011,WangYZ2012,Hirth2013,Wang20131646,Demkowicz2013,Wang201440,
Demkowicz2015234,DemkowiczCMS2016,Vattre2017,Vattre2018,Shao2018,CD2018}.
These properties  strongly depend on the characteristics of the  dislocation networks of the semicoherent interfaces.
%which depends on the interface crystallographic characters such as the misorientation and interface plane orientation.
%which gives rise to the study of semicoherent interface structures to have a better understanding and utilization of the effects of interface structure to the material properties.
%To understand and control the properties of polycrystals, it is desirable to study the interface structures.
%The prediction of material properties from the interface structure is a long-standing challenge, and depends on the .
%Atomistic simulations with explicitly calculation of each atom in the interface is a straightforward method to study the interface structure. However, this method is not always efficient.

%In a semicoherent interface, the lattice structures of adjacent crystals on its two sides are different, which leads to the incompatibilities between two lattices across the interface. To remove the incompatibilities, the misfit dislocations are included in the interface.

A semicoherent interface is composed of a network of misfitting dislocations and coherent regions separated by these dislocations. The equilibrium dislocation structure on a semicoherent interface is governed by the Frank-Bilby equation~\cite{Frank1950,Bilby1955}, which  determines the net Burgers vector content $\mathbf{B}(\mathbf{p})$ crossing a probe vector $\mathbf{p}$ on the interface. The Frank-Bilby equation strongly depends on the reference state and the possible Burgers vectors defined associated with it, on which there have been some in-depth discussions in the literature, e.g., \cite{Hirth2013,Wang20131646,Demkowicz2013,Wang201440,Demkowicz2015234,DemkowiczCMS2016,Vattre2017,Vattre2018}.

However, even though the reference state and all the associated possible Burgers vectors on a semicoherent interface are determined, the Frank-Bilby equation is still not able to give a unique dislocation structure.
See the example discussed at the end of Sec.~\ref{sec:setting}.
Wang {\it et al.} \cite{Wang20131646,Wang201440} have developed an atomically informed Frank-Bilby  theory by combining the classical Frank-Bilby theory and  atomistic simulations to determine the reference lattice and  interfacial dislocation structure of a heterogeneous interface. The dislocation line directions and their Burgers vectors in the dislocation structure in the Frank-Bilby equation are informed by atomistic simulation results.
%They obtain the dislocation structure by solving the Frank-Bilby theory while satisfying the relaxed structure in the atomistic simulation.
Vattre and Demkowicz, Abdolrahim and Demkowicz \cite{Demkowicz2013,Demkowicz2015234,DemkowiczCMS2016} have formulated approaches to determine the reference state for the interfacial misfitting dislocation arrays,
 linking the Frank–Bilby equation and anisotropic elasticity theory under the condition of vanishing far-field stresses. They considered two sets of misfitting dislocations (i.e., dislocations with two possible Burgers vectors) in their theory, for which the Frank–Bilby equation is able to give a unique solution.
Generalization of this approach has been proposed by Vattre \cite{Vattre2017} to incorporate hexagonal misfitting dislocation networks with new dislocation segments with the third Burgers vector formed by dislocation reaction, from the lozenge dislocation network of two sets of dislocations solved using the method in Ref.~\cite{Demkowicz2013,Demkowicz2015234}. Further generalizations have been made by Vattre and Pan \cite{Vattre2018} for interaction and movements of various dislocations in anisotropic bicrystals with semicoherent interfaces. There are also simulations for the dislocation/disconnection structures on semicoherent interfaces of precipitates with prescribed Burgers vectors~\cite{Quek2011,WangYZ2012}.

%The interfacial dislocations at equilibrium state cancel the long-range elastic field.
%In the previous studies, the atomistic simulation with time-consuming computation, or complicated isotropic/anisotropic elasticities are included.

In this paper, we present a continuum model to obtain the dislocation structure of a semicoherent interface, given the reference state and all possible Burgers vectors. In the continuum model, the energy of the equilibrium misfitting dislocation network is minimized with respect to all possible Burgers vectors subject to the constraint of the Frank-Bilby equation. The continuum model is based on the orientation-dependent dislocation densities of the dislocation structure. Since the Frank-Bilby equation holds, the long-range elastic energy vanishes, and the energy of the heterogeneous interface consists of only the local energy of the equilibrium dislocation network, for which the continuum formulation for the local energy of dislocation arrays \cite{Zhu2014175} is used. When the dislocation network consists of straight dislocations, our continuum model gives the exact solution (i.e., exact dislocation line directions and inter-dislocation distances) of the dislocation network.
We also develop an identification method based on dislocation reactions to recover the exact dislocation network (e.g., the hexagonal  network) from the orientation-dependent dislocation densities obtained in the continuum model. This model is a generalization of the method proposed in Ref.~\cite{Zhang2017} for finding dislocation structures of low angle grain boundaries.
%Mathematically, the choice of reference lattice does not affect the calculated spacing and line directions of interface dislocations~\cite{Demkowicz2013}, in practice, one of the adjacent crystal lattices, or a ``median lattice'' has often been used as the reference lattice.

Numerically, the constrained minimization problem in our continuum model is solved by the penalty method.
We use our continuum model to study the fcc/bcc semicoherent interfaces with the Nishiyama-Wassermann (NW) and Kurdjumov-Sachs (KS) orientation relations. Comparisons with atomistic simulations  show that our continuum model can provide excellent predictions of the dislocation structures of semicoherent interfaces.

This paper is organized as follows. In Sec.~\ref{sec:setting}, we review the semicoherent interfaces and the Frank-Bilby equation. In Sec.~\ref{sec:model}, we present our continuum model for the dislocation structure. The reference lattices and possible Burgers vectors of some semicoherent interfaces are reviewed in Sec.~\ref{sec:reference_state}. In Sec.~\ref{sec:simulation}, we apply our continuum model to obtain the dislocation structures of Cu/Nb semicoherent interfaces, and compare the results with those of atomistic simulations in Ref.~\cite{Wang20131646,Wang201440}. Conclusion and discussion are made in Sec.~\ref{sec:conclusion}.

\section{Semicoherent interfaces and Frank-Bilby equation}\label{sec:setting}
\begin{figure}[htbp]
\centering
\subfigure[]{   \includegraphics[width=0.37\linewidth]{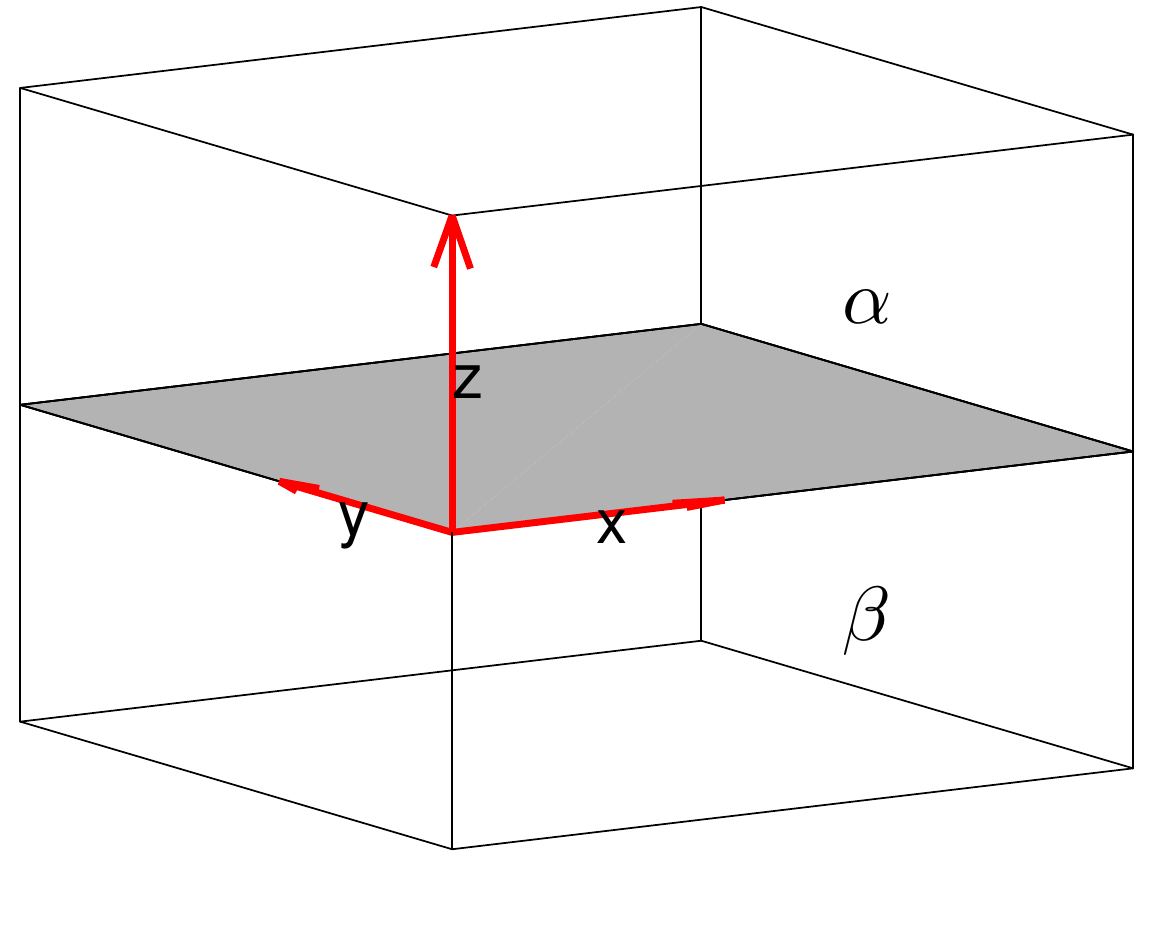}}\hspace{10mm}
\subfigure[]{    \includegraphics[width=0.35\linewidth]{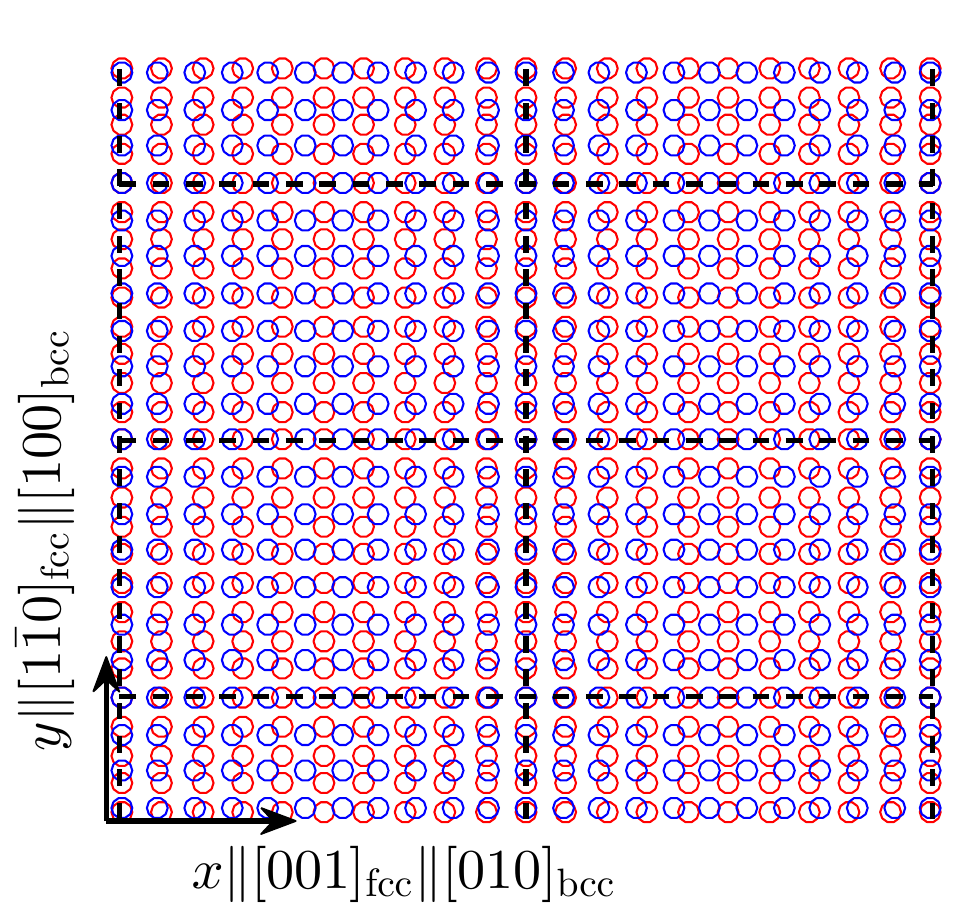}}
    \caption{(a) Illustration of the bicrystal hetero-interface. (b) The natural dichromatic pattern of the interface between fcc$(110)$/bcc$(001)$.}
    \label{fig:geometry_demo}
\end{figure}

%In this section,
We first review the semicoherent interfaces and the Frank-Bilby equation.
The geometry of a bicrystal hetero-interface is illustrated schematically in Fig.~\ref{fig:geometry_demo}(a).
Two materials (or two phases) $\alpha$ and $\beta$ with different lattice structures are joined, and a hetero-interface is formed between them.
The interface plane is set as the $xy$ plane.
When the lattice structures of the adjacent crystals are similar, and the lattice spacing difference between the unstrained adjacent crystals are relatively small, the interface usually becomes  semicoherent by forming a network of misfit dislocations on the interface, and atoms in two adjacent lattices are adjusted by additional strains or rotations.
Figure~\ref{fig:geometry_demo}(b) demonstrates a natural dichromatic pattern of the semicoherent interface between fcc$(110)$/bcc$(001)$.
%, and misfit dislocations are formed on the interface plane.
%%In general, the rotation between the adjacent lattices contains interface misorientations of tilt and twist boundaries.
%The interface is not coherent, but rather semicoherent due to the misfit dislocations formed on the interface.
%A semicoherent interface is composed of a network of misfit dislocations and coherent regions separated by misfit dislocations.

Frank and Bilby proposed a theory that provides an geometry constraint of the equilibrium dislocation structure on a semicoherent interface \cite{Frank1950,Bilby1955}.
 Given the reference lattice,
%at natural states can be related to a coherent reference lattice by uniform displacement transformation, which can be realized by the corresponding matrices $\mathbf S_{\alpha}$ and $\mathbf S_{\beta}$.
the Frank-Bilby theory determines the net Burgers vector $\mathbf{B}(\mathbf{p})$ crossing an interfacial probe vector $\mathbf{p}$ as
%Recall that the lattices $\alpha$ and $\beta$ of an interface can be related to a reference lattice by homogeneous distortion transformation matrices $S_{\alpha}$ and $S_{\beta}$.
%\begin{equation}\label{eqn:FB1}
$\mathbf{B}(\mathbf{p})=(\mathbf S_{\beta}^{-1} - \mathbf S_{\alpha}^{-1})\mathbf{p}$,
%\end{equation}
where $\mathbf S_{\alpha}^{-1}, \mathbf S_{\beta}^{-1}$ respectively are inverse matrices of the distortion transformation matrices $\mathbf S_{\alpha}, \mathbf S_{\beta}$ that map the lattice vectors from the natural unstrained lattices $\alpha$, $\beta$ to the reference lattice.

Once the reference lattice is determined, there will be finite number of possible Burgers vectors associated with it, which are the lattice vectors in the reference lattice. We denote these Burgers vectors by $\mathbf b_j$, $j=1,2,\cdots,J$.
The net Burgers vector $\mathbf{B}(\mathbf{p})$ can be expressed in the reference lattice as
%\begin{equation}\label{eqn:FB2}
$\mathbf{B}(\mathbf{p})=\sum_{j=1}^J (\mathbf N_j\cdot\mathbf{p})\mathbf{b}_j$.
%\end{equation}
Recall that in the classical dislocation model of grain boundaries and interfaces \cite{Sutton1995,Hirth2001}, the dislocation structure on a boundary/interface is described by the reciprocal vector $\mathbf N$  lying in the boundary/interface plane that is perpendicular to the dislocation and has magnitude $N=1/D$, where $D$ is the inter-dislocation distance.
The local dislocation line direction is $\bm{\xi} = (\mathbf N / N)  \times \mathbf{n}$, where $\mathbf{n}$ is the unit normal vector of the interface (which is in the $+z$ direction here). The dislocation density is $N$.
 %Using our dislocation density potential function representation, $\mathbf N=\nabla \eta$. The advantage of our dislocation density potential function representation is that it also applies to curved dislocations on curved grain boundaries while maintain the connectivity of dislocations.
For multiple arrays of dislocations on the interface, the dislocation array with Burgers vectors $\mathbf b_j$ are represented by $\mathbf N_j$, $j=1,2,\cdots,J$.
Accordingly, the {\bf Frank-Bilby equation} can be written as
\begin{flalign}
\sum_{j=1}^J \left(\mathbf N_j \cdot \mathbf{p}\right)\mathbf{b}_{j} = (\mathbf S_{\beta}^{-1} - \mathbf S_{\alpha}^{-1})\mathbf{p},
\end{flalign}
for any interfacial probe vector $\mathbf{p}$.

The Frank-Bilby equation  in general is not able to uniquely determine the dislocation structure.
For example, on the  fcc$(110)$/bcc$(001)$ interface with $[001]_{\rm fcc}\|[010]_{\rm bcc}$  and $[1\bar{1}0]_{\rm fcc}\|[100]_{\rm bcc}$, there are four possible Burgers vectors~\cite{Wang20131646}, leading to $8$ unknowns of the four vectors $\mathbf N_j=(N_{jx},N_{jy})$  for the orientations and inter-dislocation distances of the four sets of Burgers vectors; whereas there are only $4$ equations in  Frank-Bilby theory, which are not enough to determine the $8$ unknown quantities. Only when there are no more than two possible Burgers vectors in the semicoherent interface, the Frank-Bilby equation can uniquely determine a dislocation structure.

Our continuum model will be based on the given reference state.
%which determines the distortion matrices $\mathbf S_\alpha$ and $\mathbf S_\beta$ and the  possible Burgers vectors $\mathbf b_j$, $j=1,2,\cdots, J$.
In this paper, we simply adopt the median lattice (average of the two lattices) with isotropic elasticity. In practice, the median lattice or one of the adjacent lattices  have often been used as the reference lattice \cite{Frank1950,Knowles1982,Sutton1995}.  Especially, the median lattice \cite{Frank1950} is an excellent approximation of the reference lattice for symmetric and isotropic interfaces,  leading to equal partition of the elastic fields of the two crystals.
Recently, methods of determining the reference lattice in general cases such as anisotropic or unsymmetrical interfaces have also been developed \cite{Hirth2013,Wang20131646,Demkowicz2013,Wang201440,Demkowicz2015234,DemkowiczCMS2016}.

 The reference lattices and possible Burgers vectors of some hetero-interfaces will be reviewed in Sec.~\ref{sec:reference_state}.

%\subsection{Interface energy}
%We apply an energy formula derived for the low-angle grain boundary in Ref.~\cite{Zhu2014175}, where $A$ and $B$ are crystals with the same lattice structure.
%In fact, a full elastic energy of an interface $E= E_{\text{long}} + E_{\text{local}}$ was obtained in Ref.~\cite{Zhu2014175} (their Eqs.~(18)-(20), where $E_{\text{long}}$ is the long-range elastic energy and $E_{\text{local}}$ is the dislocation line energy) in terms of densities of the constituent dislocations represented by the dislocation density potential functions (see also the elastic energy expression for dislocation densities in the bulk \cite{nelson1981,rickman1997}).
%For the equilibrium planar interfaces considered here, the long-range elastic energy vanishes.
%The total energy of the interface is the local energy in terms of densities of the constituent dislocations represented by the dislocation density potential functions.
%\begin{eqnarray}
%&&E = {\displaystyle \int_S \gamma_{\rm gb}  dS},\label{eqn:gb_energy}\\
%{\rm with}&&\gamma_{\rm gb}={\displaystyle \sum_{j=1}^J}{\textstyle  \frac{\mu(b^{(j)})^2}{4\pi(1-\nu)}\!\left(1-\nu\frac{(\mathbf N_j\! \times \!\mathbf{n}   \cdot  \mathbf{b}^{(j)})^2}{(b^{(j)})^2 {\|\nabla \eta_j\|}^2}\right)\!N_j \log\! \frac{1}{r_g\sqrt{N_j^2+\epsilon}}},\label{eqn:gb_density}
%\end{eqnarray}

\section{Continuum model}\label{sec:model}

Now we present a continuum model to obtain the dislocation structure of the semicoherent hetero-interface with the given reference state (or equivalently, the distortion transformation matrices $\mathbf S_\alpha$ and $\mathbf S_\beta$) and all possible Burgers vectors ($\mathbf b_j$, $j=1,2,\cdots, J$). %from the interface energy based on densities of the constituent dislocations.
Since Frank-Bilby equation is not sufficient to uniquely determined the dislocation structure on the interface, we identify the equilibrium dislocation structure by minimizing the local energy associated with the constituent dislocations of the interface subject to the constraint of
the Frank-Bilby equation. That is, we solve the following {\bf constrained energy minimization problem} for the dislocation structure:
\begin{eqnarray}
\text{minimize}
&& E = \int_S \gamma  dS,\label{eqn:gb_energy}\\
{\rm with}&&\gamma = \sum_{j=1}^J  \frac{\mu b_{j}^2}{4\pi(1-\nu)}\!\left(1-\nu\frac{(\mathbf N_j \times \mathbf{n}   \cdot  \mathbf{b}_{j})^2}{b_{j}^2 N_j^2}\right)\!N_j \log\! \frac{1}{r_g\sqrt{N_j^2+\epsilon}},\label{eqn:gb_density}\\
\text{subject to}
&&\mathbf{h}=  \sum_{j=1}^J \left(\mathbf N_j \cdot \mathbf{p}\right)\mathbf{b}_{j} - (\mathbf S_{\beta}^{-1} - \mathbf S_{\alpha}^{-1})\mathbf{p} =\mathbf 0.\label{eqn:frank0}
\end{eqnarray}
Here  $S$ is a periodic cell on the interface plane, $\gamma$ is the interface energy density,
 $\mu$ is the shear modulus,  $\nu$ is the Poisson ratio,  $b_{j}$ is the length of the $j$-th Burgers vector, $r_g$  is a parameter associated with the dislocation core size,
 $\epsilon$ is some small positive regularization parameter to avoid the numerical singularity when $N_j=0$. The constraint in Eq.~\eqref{eqn:frank0} is the Frank-Bilby equation, which holds for any  vector $\mathbf{p}$ on the interface.

The interface energy $\gamma$ in Eq.~\eqref{eqn:gb_density} is based on the local energy of dislocation arrays in terms of dislocation densities~\cite{Zhu2014175}. Since Frank-Bilby equation is equivalent to  cancellation of the long-range elastic field~\cite{Frank1950,Bilby1955,Hirth2001,Sutton1995}, there is only local energy of the constituent dislocations on the interface when Frank-Bilby equation in Eq.~\eqref{eqn:frank0} holds, as that for the grain boundaries in homogeneous materials~\cite{Hirth2001,Sutton1995}.
The elastic constants in $\gamma$ in Eq.~\eqref{eqn:gb_density} can be chosen as the averages of those of the two materials: e.g., $\mu=(\mu_\alpha+\mu_\beta)/2$, $\nu=(\nu_\alpha+\nu_\beta)/2$.  More accurate values can be adopted if necessary.

%we remove the long-range elastic energy in the total energy formula (thus the total energy is the local energy, i.e., $E= E_{\text{local}}$ as given by Eqs.~\eqref{eqn:gb_energy} and \eqref{eqn:gb_density}) while add the constraint of Frank-Bilby equation.
%%As discussed above, the Frank-Bilby equation in general is not able to uniquely determine the dislocation structure, we minimize the interface energy over all the solutions that satisfy the Frank-Bilby equation, leading to the formulation of the constrained minimization problem presented above.

Note that the energy $\gamma$ in Eq.~\eqref{eqn:gb_density} is based on densities of dislocations. When all the constituent dislocations on the interface are straight, the obtained vectors $\mathbf N_j$'s give the exact dislocation structure. When the dislocation network consists of disconnected dislocation segments, e.g., the hexagonal network, our continuum model gives the line directions and densities of these dislocations (line direction $\mathbf N_j/N_j$ and density $N_j$); and in this case, we will present a method to recover the exact hexagonal network  from the obtained dislocation densities and line directions; see the end of this section.

The constraint of the Frank-Bilby formula in Eq.~\eqref{eqn:frank0} holds for any probe vector $\mathbf p$ if and only if it holds for the two basis vectors of the $xy$ plane: $\mathbf{p}=\mathbf{p_1}=(1,0)$ and $\mathbf{p}=\mathbf{p_2}=(0,1)$, i.e.,
 $\tilde{\mathbf h}=(\mathbf h_1,\mathbf h_2)^T=\mathbf 0$ with $\mathbf h_1$ and $\mathbf h_2$ being the Frank-Bilby formula in Eq.~\eqref{eqn:frank0} when the probe vector $\mathbf p$ is set to be $\mathbf p_1$ and $\mathbf p_2$, respectively.

 In addition to the misfit, the Frank-Bilby equation in Eq.~\eqref{eqn:frank0}  may also include further twist and/or tilt of the two crystals $\alpha$ and $\beta$ through the transformation matrices $\mathbf S_\alpha$ and  $\mathbf S_\beta$ \cite{Hirth2013,Wang20131646,Demkowicz2013,Wang201440,Demkowicz2015234,DemkowiczCMS2016,Vattre2017,Vattre2018}.
When there are rotations around an axis perpendicular to the interface (i.e., twist), assuming that the rotation angles of the natural lattices of the two materials and the reference lattice are $\theta_{\alpha}$, $\theta_{\beta}$ and $\theta$, respectively, the Burgers vectors  in the rotated reference lattice can be calculated using the rotation matrix $\mathbf R_\theta$ as $\mathbf b_i^{\rm R} = \mathbf R_\theta \mathbf b_i$, and the distortion transformation matrices are
$\mathbf S_{\alpha}^{\rm R} = \mathbf R_\theta \mathbf S_{\alpha} \mathbf R_{\alpha}^{\rm T}$ and
$\mathbf S_{\beta} ^{\rm R} = \mathbf R_\theta \mathbf S_{\beta} \mathbf R_{\beta}^{\rm T}$, where $\mathbf b_i$, $\mathbf S_{\alpha}$ and $\mathbf S_{\beta}$ are those in the un-rotated state.

{\bf Remarks}:

1. The interface energy formula in Eq.~\eqref{eqn:gb_density} can also be generalized to include elastic anisotropy based on the energy of straight dislocations with appropriate pre-logarithmic energy coefficients \cite{Hirth2001}.

2. We adopt the local energy of the constituent dislocations of the interface for a simple form and efficient calculation of the continuum model. In principle, the local energy in Eq.~\eqref{eqn:gb_density} can be replaced by the full elastic energy with elastic anisotropy and/or different elastic constants in the two materials~\cite{Srolovitz1998,Demkowicz2013,Demkowicz2015234,DemkowiczCMS2016,Vattre2017,Vattre2018} for more accurate results.

Numerically, the constrained minimization problem can be solved by the penalty method \cite{Chong2013},
 in which it is approximated by the following {\bf unconstrained minimization problem}:

\begin{equation}\label{penalty}
\text{minimize } Q = \int_S \left(\gamma   +\frac{\alpha_p}{2} \|\tilde{\mathbf h}\|^2\right) dS,
\end{equation}
where $\alpha_p >0$ with large value is the penalty parameter.
It has been shown that as the penalty parameter $\alpha_p\rightarrow+\infty$, the solution of this unconstrained minimization problem converges to the solution of the constrained minimization problem \cite{Chong2013}. (Other method such as the augmented Lagrangian method can also be used  to solve this constrained minimization problem~\cite{Chong2013}.)

This unconstrained problem is still very challenging to solve due to the nonconvexity of the interface energy. We make a further simplification by considering uniform distributions of straight dislocations on the interface. In this case, each $\mathbf N_j=(N_{jx},N_{jy})$ is a constant vector, and the problem is reduced to minimize $q=\gamma+\alpha_p \|\tilde{\mathbf h}\|^2/2$.

% in which  the probe vector $\mathbf{p}$  in $\mathbf h$ is set to be the basis vectors in the $xy$ plane $\mathbf{p_1}=(1,0)$ and $\mathbf{p_2}=(0,1)$.
%
%Assume that in the current coordinate system where the grain boundary plane is the $xy$ plane and its normal direction is the $z$ direction, the Burgers vectors are  $\mathbf{b}^{(j)}=(s_{j1},s_{j2},s_{j3})b^{(j)}$, $j=1,2,\cdots,J$, and the rotation axis is $\mathbf{a}=(a_1,a_2,a_3)$.
%The Frank-Bilby formula in Eq.~\eqref{eqn:frank0} holds if and only if it holds for the two basis vectors of the $xy$ plane: $\mathbf{p}=\mathbf{p_1}=(1,0)$ and $\mathbf{p}=\mathbf{p_2}=(0,1)$.
This unconstrained problem can be solved by gradient minimization
with respect to variables $N_{jx}$,  $N_{jy}$, $j=1,2,\cdots,J$, which
leads to the following evolution equations with an artificial time:
\begin{eqnarray}\label{eqn:evolution0}
(N_{jx})_t =-\left( \frac{\partial \gamma}{\partial N_{jx}} + \alpha_p\frac{\partial c}{\partial N_{jx}}\right),\ \ \
(N_{jy})_t =-\left( \frac{\partial \gamma}{\partial N_{jy}} + \alpha_p\frac{\partial c}{\partial N_{jy}}\right),
\end{eqnarray}
for $j=1,2,\cdots,J$, where  $c=\|\tilde{\mathbf h}\|^2/2$,
and $\tilde{\mathbf h}=(h_1,h_2,h_3,h_4)^T$ with
 $h_1 = \sum_{j=1}^J b_{jx} N_{jx} - (\mathbf S_{\beta}^{-1}[1,1] - \mathbf S_{{\alpha}}^{-1}[1,1])$,
 $h_2 = \sum_{j=1}^J b_{jy} N_{jx} - (\mathbf S_{\beta}^{-1}[2,1] - \mathbf S_{{\alpha}}^{-1}[2,1])$,
 $h_3 = \sum_{j=1}^J b_{jx} N_{jy} - (\mathbf S_{\beta}^{-1}[1,2] - \mathbf S_{{\alpha}}^{-1}[1,2])$,
 $h_4 = \sum_{j=1}^J b_{jy} N_{jy} - (\mathbf S_{\beta}^{-1}[2,2] - \mathbf S_{{\alpha}}^{-1}[2,2])$.
Note that when tilt of the two crystals $\alpha$ and $\beta$ is also considered, in the Frank-Bilby equation in Eq.~\eqref{eqn:frank0}, $\mathbf S_\alpha$ and $\mathbf S_\beta$ will be $3\times 3$ matrices, $\mathbf b_j$, $\mathbf p$ and $\mathbf h$ will be vectors in three dimensions, and $\tilde{\mathbf h}$ will be a $6\times 1$ vector.

{\bf Identification of dislocation structure from dislocation densities.}
Now we present a method that recovers the exact dislocation structure based on the densities and orientations of the constituent dislocations obtained in our continuum model.  As mentioned above, when all the constituent dislocations on the interface are straight, the obtained vectors $\mathbf N_j$'s give the exact dislocation structure directly.
In a general dislocation structure, due to dislocation reactions, the dislocations may not necessarily be continuous straight lines, and they may form hexagons (not necessarily regular) with disconnected dislocation segments; see  Fig.~\ref{fig:Hex_CDP}(b) for an example.
%In the modified continuum model presented above, we have successfully identified dislocation reactions in the dislocation structure of a grain boundary. This enables us to draw the exact dislocation structure based on the results of the continuum model. The method is described below.
In the identification method, we calculate the exact length and orientation of each dislocation segment in the hexagonal network based on the dislocation densities and orientations obtained by the continuum model.

\begin{figure}[htbp]
	\centering
	\includegraphics[width=3.3in]{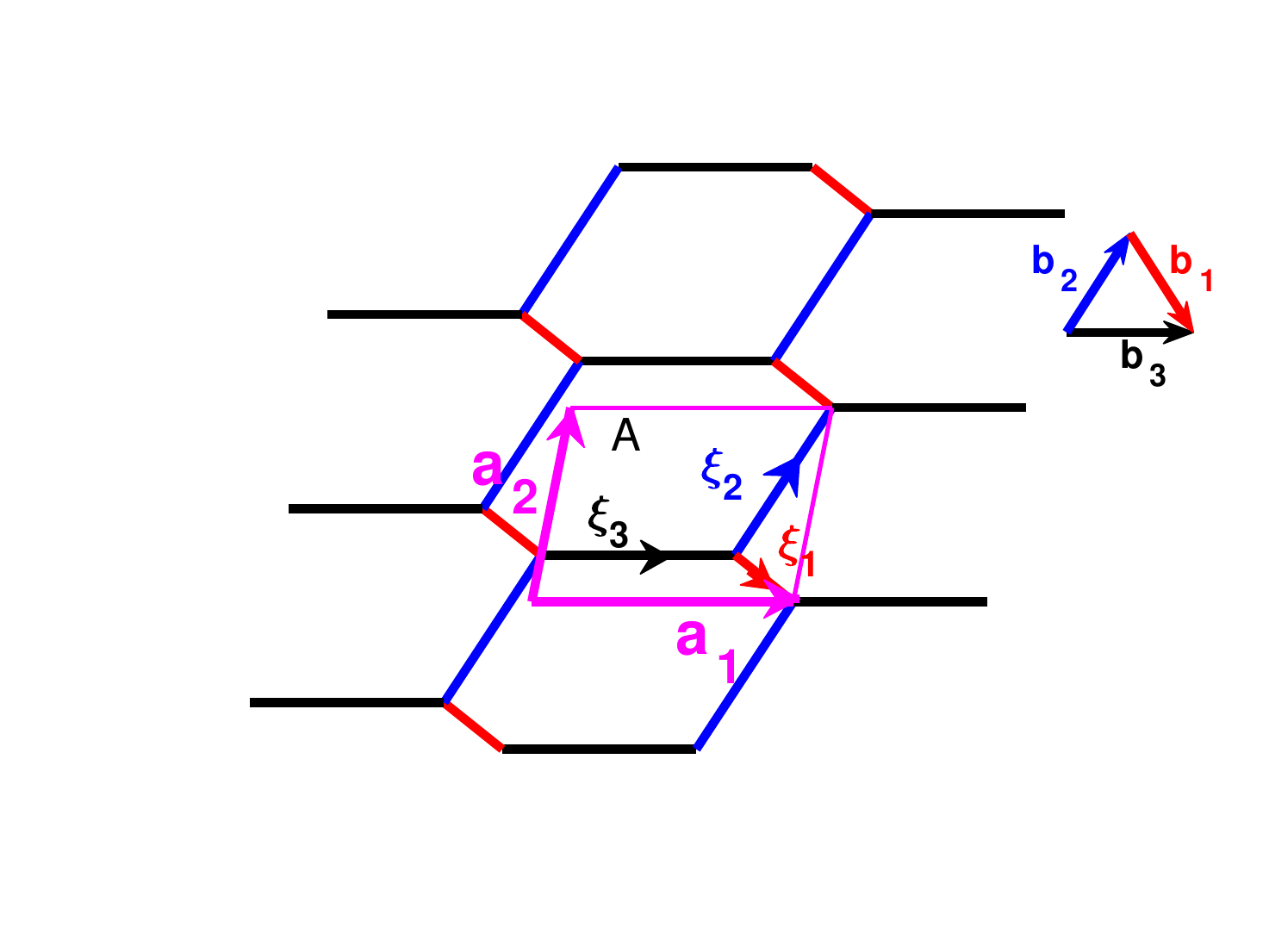}
	\caption{A hexagonal dislocation structure that consists of dislocations with Burgers vectors $\mathbf b_1$, $\mathbf b_2$ and $\mathbf b_3$, whose line directions are  $\pmb\xi_1$, $\pmb\xi_2$, and $\pmb\xi_3$, respectively. Vectors $\mathbf a_1$ and $\mathbf a_2$ are the two sides of the periodic parallelogram unit cell.  The area of a unit cell is $A=\| \mathbf a_1\times\mathbf a_2\|$.   }
	\label{fig:hex}
\end{figure}

Consider a hexagonal network with dislocations of three Burgers vectors $\mathbf b_1$, $\mathbf b_2$, and $\mathbf b_3$, in which dislocations may have reactions, e.g, $\mathbf b_3=\mathbf b_1+\mathbf b_2$ and $b_2^2<b_1^2+b_2^2$; see Fig.~\ref{fig:hex}. In this hexagonal network, as we discussed before, the direction of dislocation generated by $\mathbf b_j$ is $ \pmb\xi_j = (\mathbf N_j/N_j)\times \mathbf{n}$, and the  density of these dislocations is $N_j$.
 %This means that the length of a $\mathbf b_j$-dislocation segment in a periodic parallelogram cell with area $A$ is $\|\nabla_s\eta_j\|A$.
 Consider a periodic parallelogram cell as shown in Fig.~\ref{fig:hex}.  It can be calculated that the $\mathbf b_1$-, $\mathbf b_2$-, and $\mathbf b_3$-dislocation segments in the parallelogram, written in the vector form, are
 $\mathbf l_1=A\mathbf N_1\times \mathbf n$,  $\mathbf l_2=A\mathbf N_2\times \mathbf n$,   $\mathbf l_3=A\mathbf N_3\times \mathbf n$,
 respectively,  where $\mathbf n$ is the normal vector of the interface, and $A$ is the area of the periodic parallelogram cell.
 %and its area is $A=\|\mathbf l_1\times \mathbf l_2\|+\|\mathbf l_2\times \mathbf l_3\|+\|\mathbf l_3\times \mathbf l_1\|$.
  %it can be calculated that its area is $A=1/(\| \mathbf N_1\times \mathbf N_2 \| +\|\mathbf N_2\times \mathbf N_3 \| +\|\mathbf N_3\times \mathbf N_1\|)$.
%Using this result,
Using these results, it can be solved that in a periodic parallelogram cell,
the length of each dislocation segment is
\begin{equation}\label{eqn:segmentlength}
l_j=\frac{N_j}{\| \mathbf N_1\times \mathbf N_2 \| +\|\mathbf N_2\times \mathbf N_3 \| +\|\mathbf N_3\times \mathbf N_1\| }, \ \ j=1,2,3,
\end{equation}
 Using this formula of length of each dislocation segment $l_j$  and
its direction $\pmb\xi_j= \left(\mathbf N_j/N_j\right)  \times \mathbf{n}$,  we can draw the exact hexagonal network structure
 based on the dislocation densities and orientations  represented by $\{\mathbf N_j\}$ in the continuum model.

\section{Reference lattice and possible Burgers vectors}\label{sec:reference_state}

In this section, we briefly review the  reference lattice and possible Burgers vectors for the dislocation network on a semicoherent interface. The distortion transformation matrices $\mathbf S_{\alpha}$ and $\mathbf S_{\beta}$ are defined as the matrices mapping from the dichromatic patterns  of the two lattices ${\alpha}$ and ${\beta}$ to the reference lattice, and the possible Burgers vectors ar the lattice vectors in the reference lattice \cite{Sutton1995}.
In practice, the median lattice (average of the two lattices) or one of the adjacent lattices  have often been used as the reference lattice \cite{Frank1950,Knowles1982,Sutton1995}.  Especially, the median lattice \cite{Frank1950} is an excellent approximation of the reference lattice for symmetric and isotropic interfaces,  leading to equal partition of the elastic fields of the two crystals.
Recently, methods of determining the reference lattice in general cases such as anisotropic or unsymmetrical interfaces have also been developed \cite{Hirth2013,Wang20131646,Demkowicz2013,Wang201440,Demkowicz2015234,DemkowiczCMS2016}.

%Our continuum model presented in the previous section depends on the given reference state and the associated possible Burgers vectors. In the numerical simulations presented in this paper, we simply adopt the median lattice with isotropic elasticity. These simulations show that our continuum model works well for the dislocation structures on the Cu$(110)$/Nb$(001)$ and Cu$(111)$/Nb$(110)$ interfaces that have been studied using atomistic simulations and the atomistically informed Frank-Bilby theory \cite{Wang20131646,Wang201440}.

%The reference lattice pattern depends on the dichromatic patterns of the adjacent crystals of the interface.

\begin{figure}[!h]
\centering
\subfigure[]{  \includegraphics[width=0.35\linewidth]{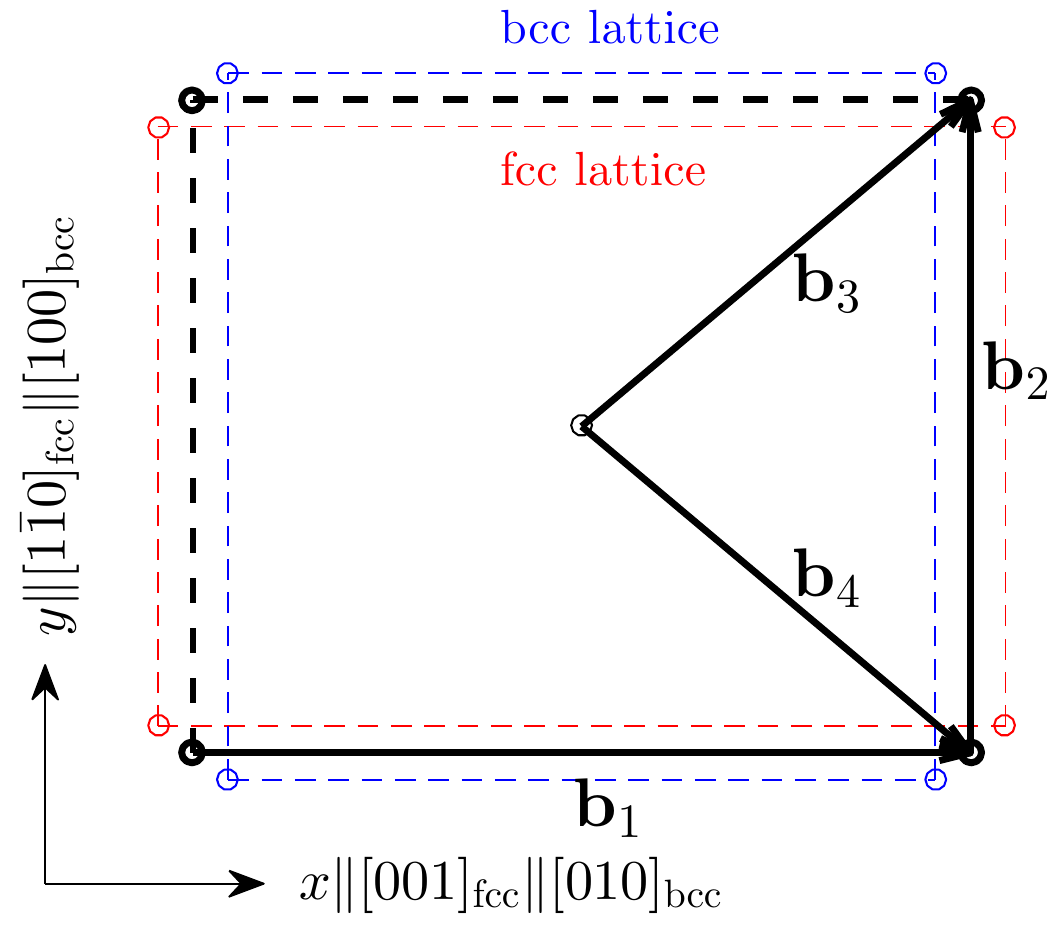}}\hspace{10mm}
\subfigure[]{  \includegraphics[width=0.35\linewidth]{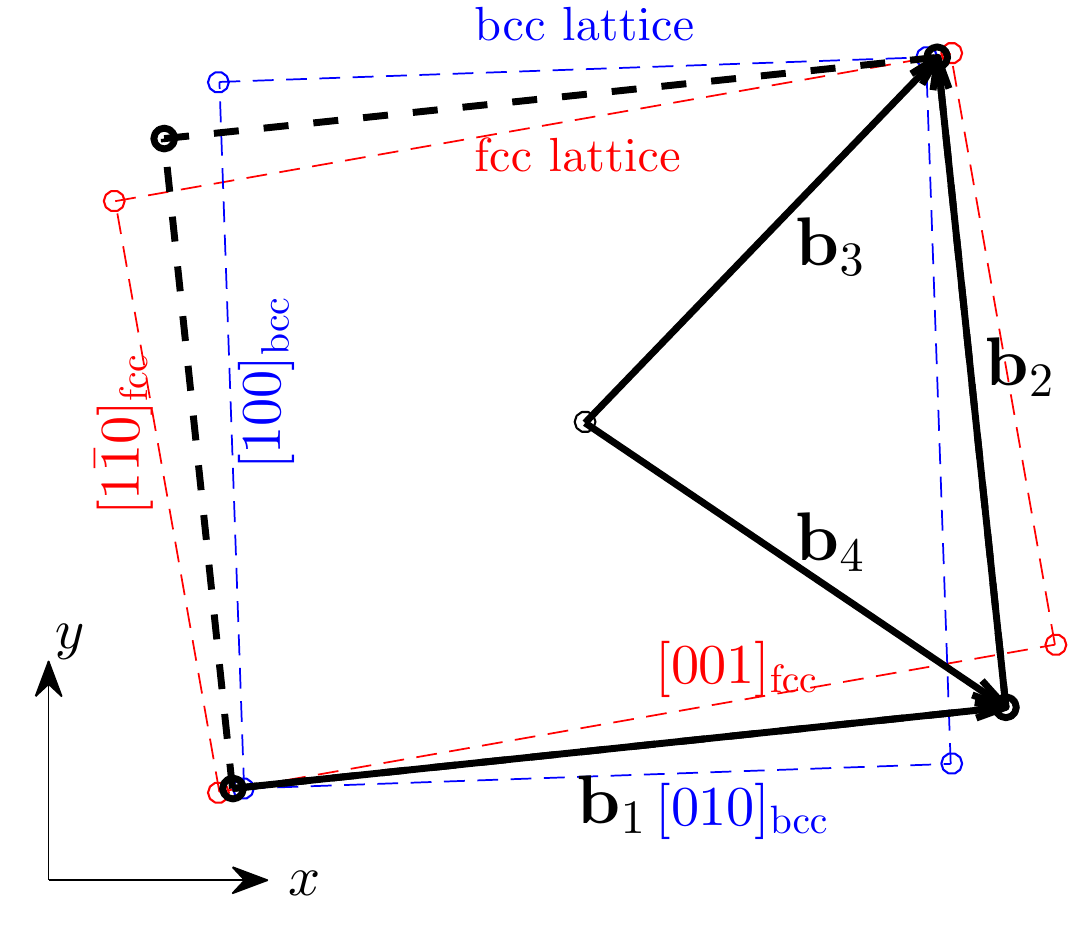}}
\subfigure[]{  \includegraphics[width=0.32\linewidth]{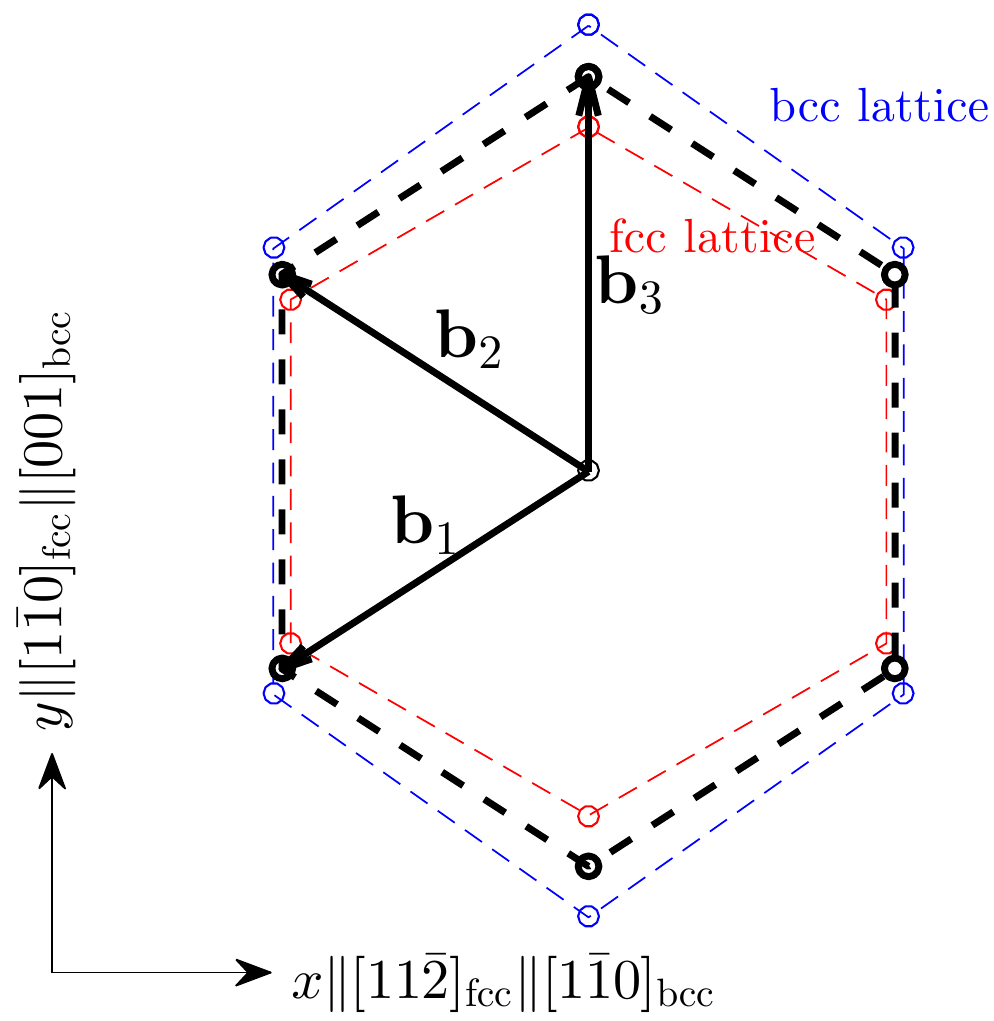}}\hspace{10mm}
\subfigure[]{  \includegraphics[width=0.32\linewidth]{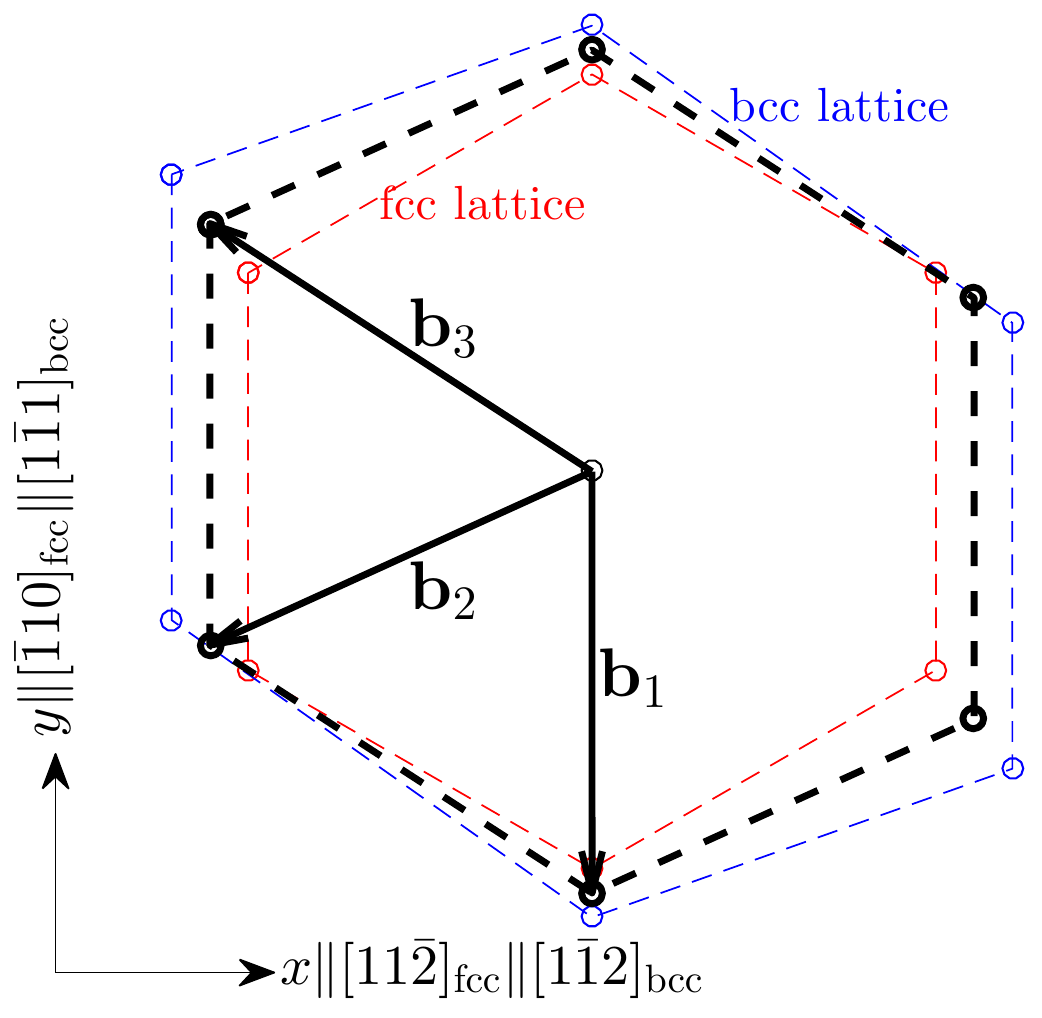}}
    \caption{Geometry of the  dichromatic patterns, reference states and possible Burgers vectors of some classical semicoherent interfaces  between fcc and bcc lattices.
    The natural dichromatic patterns of the fcc and bcc lattices and the reference lattice (the median lattice) are polygons with  red, blue, and black dashed lines, respectively. The possible Burgers vectors defined in the reference lattice are shown by black arrows.
    (a) Interface with the orientation relationship $(110)_{\rm fcc}\|(001)_{\rm bcc}$, $[001]_{\rm fcc}\|[010]_{\rm bcc}$, and $[1\bar{1}0]_{\rm fcc}\|[100]_{\rm bcc}$.
The natural dichromatic patterns  are rectangles.
(b) Interface with the orientation relationship of that in (a) with further rotations around an axis perpendicular to the interface.
The reference lattice and possible Burgers vectors can all be obtained by rotations from those in  (a).
(c) Interface with the NW orientation relationship: $(111)_{\rm fcc}\|(110)_{\rm bcc}$, $[11\bar{2}]_{\rm fcc}\|[1\bar{1}0]_{\rm bcc}$, and $[1\bar{1}0]_{\rm fcc}\|[001]_{\rm bcc}$. The natural dichromatic patterns are hexagons.
(d) Interface with the KS orientation relationship: $(111)_{\rm fcc}\|(110)_{\rm bcc}$, $[11\bar{2}]_{\rm fcc}\|[1\bar{1}2]_{\rm bcc}$, and $[\bar{1}10]_{\rm fcc}\|[1\bar{1}1]_{\rm bcc}$.
The fcc and bcc lattices, reference lattice patterns and Burgers vectors  can be obtained by rotating the corresponding ones in the NW orientation relationship in (c) around an axis perpendicular to the interface. }
    \label{fig:CDP_demo}
\end{figure}

Figure~\ref{fig:CDP_demo} shows the  reference lattices and possible Burgers vectors of several semicoherent interfaces between fcc and bcc lattices that have been studied recently in the literature \cite{Hirth2013,Wang20131646,Demkowicz2013,Wang201440,Demkowicz2015234,DemkowiczCMS2016,Vattre2017,Vattre2018,Shao2018}.
In Fig.~\ref{fig:CDP_demo}(a), the interface orientation relationships are $(110)_{\rm fcc}\|(001)_{\rm bcc}$, $[001]_{\rm fcc}\|[010]_{\rm bcc}$ (in $x$-direction) and $[1\bar{1}0]_{\rm fcc}\|[100]_{\rm bcc}$ (in $y$-direction).
The natural dichromatic patterns  of the fcc and bcc lattices are rectangles with parallel sides, as shown by the red and blue dashed lines, respectively, in Fig.~\ref{fig:CDP_demo}(a).
The coherent dichromatic pattern of the reference lattice is also a rectangle pattern in between the fcc and bcc natural dichromatic patterns, as shown by the black dashed lines in Fig.~\ref{fig:CDP_demo}(a).
When the median lattice (the average of the fcc and bcc natural dichromatic patterns) is adopted as the reference lattice, the possible Burgers vectors of dislocations on the semicoherent interface are the lattice vectors of the  reference lattice:
\begin{flalign}\label{eqn:BV_rectangle}
&\textstyle{\mathbf b_1 = \left(\frac{1}{2}a_{\rm fcc}+\frac{1}{2}a_{\rm bcc}, 0\right),
\mathbf b_2 = \left(0, \frac{\sqrt{2}}{4}a_{\rm fcc}+\frac{1}{2}a_{\rm bcc}\right)},\nonumber\\
&\textstyle{\mathbf b_3 = \left(\frac{1}{4}a_{\rm fcc}+\frac{1}{4}a_{\rm bcc}, \frac{\sqrt{2}}{8}a_{\rm fcc}+\frac{1}{4}a_{\rm bcc}\right),
\mathbf b_4 = \left(\frac{1}{4}a_{\rm fcc}+\frac{1}{4}a_{\rm bcc}, -\left(\frac{\sqrt{2}}{8}a_{\rm fcc}+\frac{1}{4}a_{\rm bcc}\right)\right)},
\end{flalign}
and the distortion transformation matrices mapping from the fcc and bcc lattices  to the reference lattice are
\begin{flalign}\label{eqn:S_rectangle_CDP}
\mathbf S_{\alpha} =
\begin{pmatrix}
\frac{a_{\rm fcc}+a_{\rm bcc}}{2a_{\rm fcc}} & 0 \\
0 & \frac{\frac{\sqrt{2}}{2}a_{\rm fcc}+a_{\rm bcc}}{\sqrt{2}a_{\rm fcc}}
\end{pmatrix},\quad
\mathbf S_{\beta} =
\begin{pmatrix}
\frac{a_{\rm fcc}+a_{\rm bcc}}{2a_{\rm bcc}} & 0 \\
0 & \frac{\frac{\sqrt{2}}{2}a_{\rm fcc}+a_{\rm bcc}}{2a_{\rm bcc}}
\end{pmatrix},
\end{flalign}
where $a_{\rm fcc}$ and $a_{\rm bcc}$ are the lattice constants of the fcc and bcc lattices.

Figure~\ref{fig:CDP_demo}(b) shows the orientation relationship of the fcc$(110)$/bcc$(001)$ interface in Fig.~\ref{fig:CDP_demo}(a) with further rotations around an axis perpendicular to the interface.
%When there are rotations of the fcc and bcc lattices in Fig.~\ref{fig:CDP_demo}(a) around an axis perpendicular to the $xy$ plane, the rotation CDP (RCDP) is demonstrated in Fig.~\ref{fig:CDP_demo}(b).
The rotation angles of the natural fcc, bcc lattices and the reference lattice are $\theta_{\alpha}$, $\theta_{\beta}$ and $\theta$, respectively.
The Burgers vectors  in the rotated reference lattice can be calculated by multiplying the Burgers vectors in the un-rotated reference lattice by the rotation matrix $\mathbf R_\theta$, i.e.,
\begin{flalign}\label{eqn:BV_rotate}
\mathbf b_i^{\rm R} = \mathbf R_\theta \mathbf b_i,
\text{ with }
\mathbf  R_\theta =
\begin{pmatrix}
\cos\theta & -\sin\theta \\
\sin\theta & \cos\theta
\end{pmatrix}.
\end{flalign}
%where $\theta$ is the rotation angle of the reference lattice from the NW OR to KS OR in the counterclockwise direction.
%We denote the rotation angle of the $\alpha$ and $\beta$ lattices are respectively $\theta_{\alpha}$ and $\theta_{\beta}$.
The distortion transformation matrices are
\begin{flalign}
&\mathbf S_{\alpha}^{\rm R} = \mathbf R_\theta \mathbf S_{\alpha} \mathbf R_{\alpha}^{\rm T}, \quad
\mathbf S_{\beta} ^{\rm R} = \mathbf R_\theta \mathbf S_{\beta} \mathbf R_{\beta}^{\rm T}.\label{eqn:S_RCDP}
\end{flalign}

%\begin{flalign}
%\mathbf S_{\alpha} =
%\begin{pmatrix}
%\frac{(a_{\rm fcc}+a_{\rm bcc})\cos\theta\cos\theta_{\alpha} + (a_{\rm fcc}+\sqrt{2}a_{\rm bcc})\sin\theta\sin\theta_{\alpha}}{2a_{\rm fcc}} &
%\frac{(a_{\rm fcc}+a_{\rm bcc})\cos\theta\sin\theta_{\alpha} - (a_{\rm fcc}+\sqrt{2}a_{\rm bcc})\sin\theta\cos\theta_{\alpha}}{2a_{\rm fcc}} \\
%\frac{(a_{\rm fcc}+a_{\rm bcc})\sin\theta\cos\theta_{\alpha} - (a_{\rm fcc}+\sqrt{2}a_{\rm bcc})\cos\theta\sin\theta_{\alpha}}{2a_{\rm fcc}} &
%\frac{(a_{\rm fcc}+a_{\rm bcc})\sin\theta\sin\theta_{\alpha} + (a_{\rm fcc}+\sqrt{2}a_{\rm bcc})\cos\theta\cos\theta_{\alpha}}{2a_{\rm fcc}}
%\end{pmatrix},\\
%\mathbf S_{\beta} =
%\begin{pmatrix}
%\frac{(a_{\rm fcc}+a_{\rm bcc})\cos\theta\cos\theta_{\beta} + (\frac{\sqrt{2}}{2}a_{\rm fcc}+a_{\rm bcc})\sin\theta\sin\theta_{\beta}}{2a_{\rm bcc}} &
%\frac{(a_{\rm fcc}+a_{\rm bcc})\cos\theta\sin\theta_{\beta} - (\frac{\sqrt{2}}{2}a_{\rm fcc}+a_{\rm bcc})\sin\theta\cos\theta_{\beta}}{2a_{\rm bcc}} \\
%\frac{(a_{\rm fcc}+a_{\rm bcc})\sin\theta\cos\theta_{\beta} - (\frac{\sqrt{2}}{2}a_{\rm fcc}+a_{\rm bcc})\cos\theta\sin\theta_{\beta}}{2a_{\rm bcc}} &
%\frac{(a_{\rm fcc}+a_{\rm bcc})\sin\theta\sin\theta_{\beta} + (\frac{\sqrt{2}}{2}a_{\rm fcc}+a_{\rm bcc})\cos\theta\cos\theta_{\beta}}{2a_{\rm bcc}}
%\end{pmatrix}.
%\end{flalign}

Figure~\ref{fig:CDP_demo}(c) demonstrates the classical Nishiyama-Wassermann (NW) orientation relationship: $(111)_{\rm fcc}\|(110)_{\rm bcc}$, $[11\bar{2}]_{\rm fcc}\|[1\bar{1}0]_{\rm bcc}$ (in $x$-direction), and $[1\bar{1}0]_{\rm fcc}\|[001]_{\rm bcc}$ (in $y$-direction).
The natural dichromatic patterns   of fcc and bcc lattices are hexagonal patterns,
%(red and blue dashed lines respectively in Fig.~\ref{fig:CDP_demo}(c)),
and the coherent dichromatic pattern of the reference lattice, which is the median lattice between the fcc and bcc lattices,  is also a hexagonal pattern.
%(black dashed lines in Fig.~\ref{fig:CDP_demo}(c)).
The possible Burgers vectors defined in the reference lattice are
\begin{flalign}\label{eqn:BV_hex}
&\textstyle{\mathbf b_1 = \left(-\left(\frac{\sqrt{6}}{8} a_{\rm fcc}+\frac{\sqrt{2}}{4}a_{\rm bcc}\right), -\left(\frac{\sqrt{2}}{8}a_{\rm fcc}+\frac{1}{4}a_{\rm bcc}\right)\right)},\nonumber\\
&\textstyle{\mathbf b_2 = \left(-\left(\frac{\sqrt{6}}{8} a_{\rm fcc}+\frac{\sqrt{2}}{4}a_{\rm bcc}\right), \frac{\sqrt{2}}{8}a_{\rm fcc}+\frac{1}{4}a_{\rm bcc}\right),
\mathbf b_3 = \left(0, \frac{\sqrt{2}}{4} a_{\rm fcc}+\frac{1}{2}a_{\rm bcc}\right)},
\end{flalign}
and the distortion transformation matrices are
\begin{flalign}\label{eqn:S_hex_CDP}
\mathbf S_{\alpha} =
\begin{pmatrix}
\frac{\frac{\sqrt{6}}{4} a_{\rm fcc}+\frac{\sqrt{2}}{2}a_{\rm bcc}}{\frac{\sqrt{6}}{2} a_{\rm fcc}} & 0 \\
0 & \frac{\frac{\sqrt{2}}{2} a_{\rm fcc}+a_{\rm bcc}}{\sqrt{2}a_{\rm fcc}}
\end{pmatrix},\quad
\mathbf S_{\beta} =
\begin{pmatrix}
\frac{\frac{\sqrt{6}}{4} a_{\rm fcc}+\frac{\sqrt{2}}{2}a_{\rm bcc}}{\sqrt{2} a_{\rm bcc}} & 0 \\
0 & \frac{\frac{\sqrt{2}}{2} a_{\rm fcc}+a_{\rm bcc}}{2a_{\rm fcc}}
\end{pmatrix}.
\end{flalign}

Another classical orientation relationship of the interface jointing $(111)_{\rm fcc}\|(110)_{\rm bcc}$ is the Kurdjumov-Sachs (KS) orientation relationship with $[11\bar{2}]_{\rm fcc}\|[1\bar{1}2]_{\rm bcc}$ (in $x$-direction) and $[\bar{1}10]_{\rm fcc}\|[1\bar{1}1]_{\rm bcc}$ (in $y$-direction); see Fig.~\ref{fig:CDP_demo}(d).
It can be obtained by rotating the fcc and bcc lattices in NW orientation relationship around an axis perpendicular to the interface. The possible Burgers vectors and the distortion transformation matrices can also be obtained from those in the  NW orientation relationship by rotations using Eqs.~\eqref{eqn:BV_rotate} and \eqref{eqn:S_RCDP}.

\section{Numerical simulation}\label{sec:simulation}
We apply our continuum simulation model to obtain dislocation structures on the  semicoherent interfaces of  Cu$(110)$/Nb$(001)$  and Cu$(111)$/Nb$(110)$, which have been studied recently in the literature \cite{Hirth2013,Wang20131646,Demkowicz2013,Wang201440,Demkowicz2015234,DemkowiczCMS2016,Vattre2017,Vattre2018,Shao2018}.
The orientation relationships, reference states (the median lattices) and possible Burgers vectors of these interfaces are shown in Fig.~\ref{fig:CDP_demo}. The lattice constants of Cu and Nb are $a_{\rm Cu}=0.360$nm and $a_{\rm Nb}=0.330$nm, respectively.
%Moreover, through these simulations, we calibrate the values of $r_g$  for the  pure tilt and pure twist boundaries
%by the  available results of energies of these boundaries by using molecular dynamics (MD) simulations or other methods. These  values of $r_g$ will be extended to all grain boundary orientations by an interpolation in the next section. We use the EAM potential for Al developed by Mishin \textit{et al.}  \cite{Mishin1999} and the LAMMPS code \cite{Plimpton1995} in the MD simulations.
%Recall that in the continuum model, the locations of dislocations are described by their reciprocal vectors $\mathbf N_j=(N_{jx},N_{jy})$, $j=1,2,\cdots,J$, and in the simulations, the dislocation lines are $N_{jx} x + N_{jy} y = 0$, where $N_{jx}$ and $N_{jy}$ are constants and are obtained by solving the minimization problem in Eq.~\eqref{penalty} (that is, the evolution equations in Eqs.~\eqref{eqn:evolution0} and \eqref{eqn:evolution1}).
We start from $N_{jx}=N_{jy}=0$, $j=1,2,\cdots,J$, when performing the energy gradient minimization in Eq.~\eqref{eqn:evolution0}. We choose a large value for the penalty parameter $\alpha_p$ in Eq.~\eqref{penalty} and further increases of its value give only negligible changes in the converged dislocation structures.
We compare our results with those obtained using atomistic simulations \cite{Wang20131646,Wang201440}.
The reference lattices of these semicoherent interfaces obtained by atomistic simulations \cite{Wang20131646,Wang201440}
are very close to the median lattices  adopted in our simulations.

\subsection{Cu$(110)$/Nb$(001)$ interface with rectangle pattern}

We first consider the Cu$(110)$/Nb$(001)$ interface with the orientation relationship  as shown in Fig.~\ref{fig:CDP_demo}(a), where the natural Cu (fcc), Nb (bcc), and reference lattices have rectangle patterns.  There are four possible Burgers vectors $\mathbf b_j$, $j=1,2,3,4$. See
Eqs.~\eqref{eqn:BV_rectangle} and \eqref{eqn:S_rectangle_CDP}
for the formulas of these Burgers vectors and the distortion matrices $\mathbf S_{\alpha}$ and $\mathbf S_{\beta}$.

\begin{figure}[htbp]
\centering
%    \includegraphics[width=0.35\linewidth]{Rectangle_CDP_MD.jpg}
%\subfigure[]{  \includegraphics[width=0.32\linewidth]{Rectangle_CDP_demo.eps}}
 \includegraphics[width=0.4\linewidth]{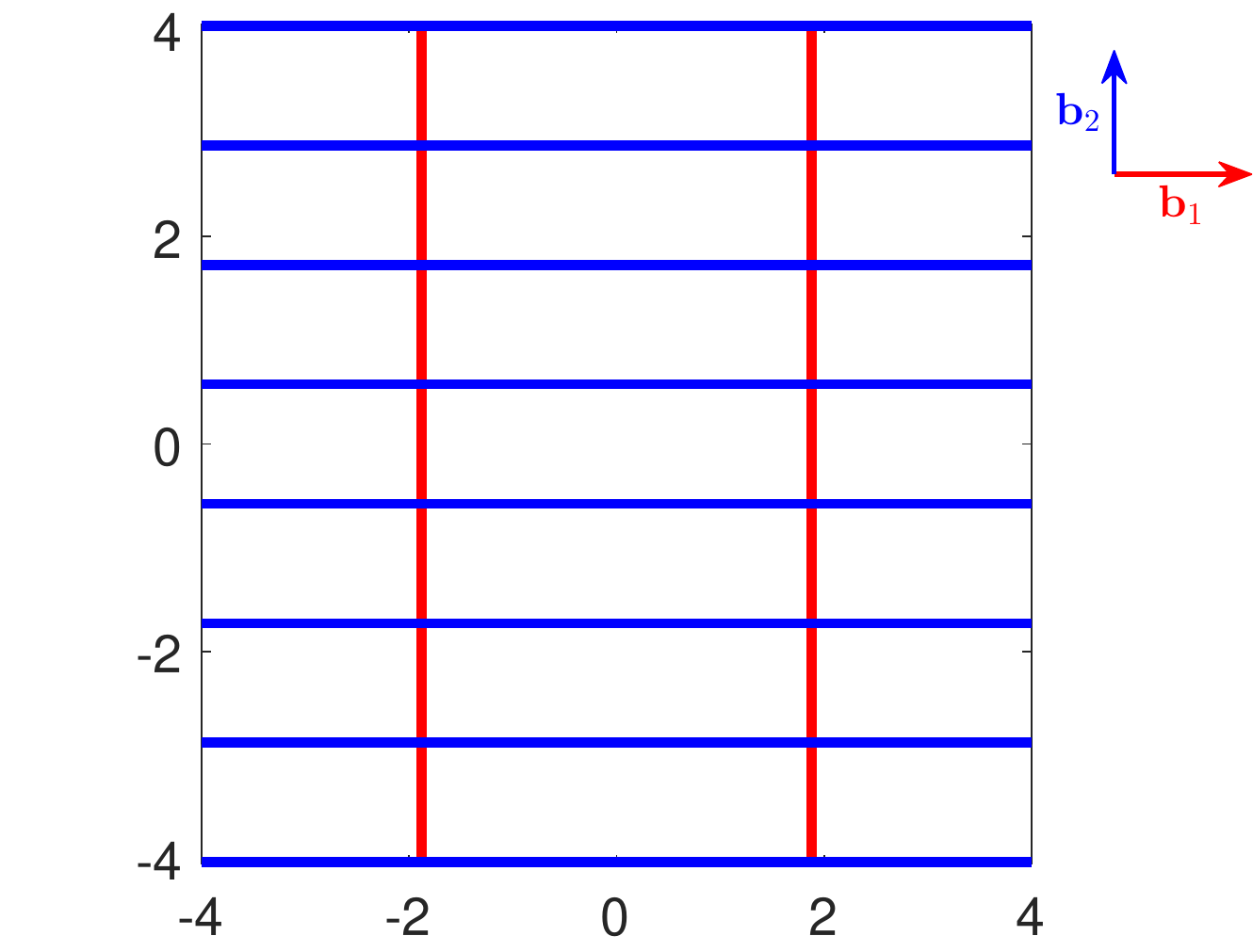}
    \caption{Dislocation structure of this Cu$(110)$/Nb$(001)$ interface  calculated using our continuum model. It is a rectangular network that consists of two arrays of dislocations with Burgers vectors $\mathbf{b}_1$ and  $\mathbf{b}_2$ (red vertical lines and blue horizontal lines, respectively).  Unit: \rm{nm}.}
    \label{fig:Rec_CDP}
\end{figure}

The dislocation structure obtained using our continuum model is shown in Fig.~\ref{fig:Rec_CDP}.
The dislocation structure is a rectangular network that consists of two arrays of dislocations with Burgers vectors $\mathbf{b}_1$, $\mathbf{b}_2$ and represented by the reciprocal vectors $\mathbf N_1=(-0.2658,0)/$nm, $\mathbf N_2=(0, 0.8681)/$nm, respectively.
%The dislocations with Burgers vector $\mathbf{b}_1$ are lying parallel to the $x\|[001]_{\rm Cu}\|[010]_{\rm Nb}$ direction, and dislocations with Burgers vector $\mathbf{b}_2$ are parallel to $y\|[1\bar{1}0]_{\rm Cu}\|[100]_{\rm Nb}$ direction.
These two arrays of dislocations are both edge dislocations, and are in the $+y$ ($[1\bar{1}0]_{\rm Cu}\|[100]_{\rm Nb}$) and $+x$ ($[001]_{\rm Cu}\|[010]_{\rm Nb}$) directions, with inter-dislocation distances  $D_1=1/{N_1} = 3.76 \rm {nm}$ and $D_2=1/{N_2} = 1.15\rm {nm}$, respectively.
%Recall that the Burgers vector $\mathbf{b}^{(1)}$ is in the $z$ direction corresponding to the $[\bar{1}10]$ direction.
Dislocations with Burgers vectors $\mathbf{b}_3$ and $\mathbf{b}_4$  do not appear in the converged dislocation structure, i.e., $\mathbf N_3, \mathbf N_4$ converge to $\mathbf 0$ in the simulation.
These results of rectangular network, dislocation line directions and inter-dislocation distances agree excellently with those obtained using atomistic simulation in Ref.~\cite{Wang20131646}, in which
 the inter-dislocation distances in the two dislocation arrays are $D_1=3.80\rm {nm}$ and $D_2=1.13\rm {nm}$.

\subsection{Cu$(111)$/Nb$(110)$ interface with NW orientation relationship}

 We then consider the Cu$(111)$/Nb$(110)$ interface with hexagonal pattern in
 the classical NW orientation relationship as shown in Fig.~\ref{fig:CDP_demo}(c).
The natural dichromatic patterns   of Cu (fcc) and Nb (bcc) lattices and the coherent dichromatic pattern of the reference lattice are hexagonal patterns.
The three possible Burgers vectors defined in the reference lattice  and the distortion matrices $\mathbf S_{\alpha}$ and $\mathbf S_{\beta}$ are given in Eqs.~\eqref{eqn:BV_hex} and
\eqref{eqn:S_hex_CDP}.

%$\mathbf b_1 = \left(-\left(\frac{\sqrt{6}}{8} a_{\rm fcc}+\frac{\sqrt{2}}{4}a_{\rm bcc}\right), -\left(\frac{\sqrt{2}}{8}a_{\rm fcc}+\frac{1}{4}a_{\rm bcc}\right)\right)$,
%$\mathbf b_2 = \left(-\left(\frac{\sqrt{6}}{8} a_{\rm fcc}+\frac{\sqrt{2}}{4}a_{\rm bcc}\right), \frac{\sqrt{2}}{8}a_{\rm fcc}+\frac{1}{4}a_{\rm bcc}\right)$, and
%$\mathbf b_3 = \left(0, \frac{\sqrt{2}}{4} a_{\rm fcc}+\frac{1}{2}a_{\rm bcc}\right)$.
%The distortion transformation matrices are\\
%$\mathbf S_{\alpha} =
%{\rm diag}\left(
%\frac{\frac{\sqrt{6}}{4} a_{\rm fcc}+\frac{\sqrt{2}}{2}a_{\rm bcc}}{\frac{\sqrt{6}}{2} a_{\rm fcc}}, \frac{\frac{\sqrt{2}}{2} a_{\rm fcc}+a_{\rm bcc}}{\sqrt{2}a_{\rm fcc}}\right)$ and
%$\mathbf S_{\beta} ={\rm diag}\left(
%\frac{\frac{\sqrt{6}}{4} a_{\rm fcc}+\frac{\sqrt{2}}{2}a_{\rm bcc}}{\sqrt{2} a_{\rm bcc}},\frac{\frac{\sqrt{2}}{2} a_{\rm fcc}+a_{\rm bcc}}{2a_{\rm fcc}}\right)$.
% Using the lattice constants of Cu and Nb, the three possible Burgers vectors are $\mathbf b_1= (-0.2267, -0.1460)$nm, $\mathbf b_2 = (-0.2267, 0.1460)$nm, $\mathbf b_3 = (0, 0.2920)$nm in the $xy$ plane shown in Fig.~\ref{fig:Hex_CDP}(a).

\begin{figure}[htbp]
\centering
%    \includegraphics[width=0.4\linewidth]{Hex_CDP_MD.jpg}
%\subfigure[]{  \includegraphics[width=0.32\linewidth]{Hex_CDP_demo.eps}}
 \subfigure[]{   \includegraphics[width=0.60\linewidth]{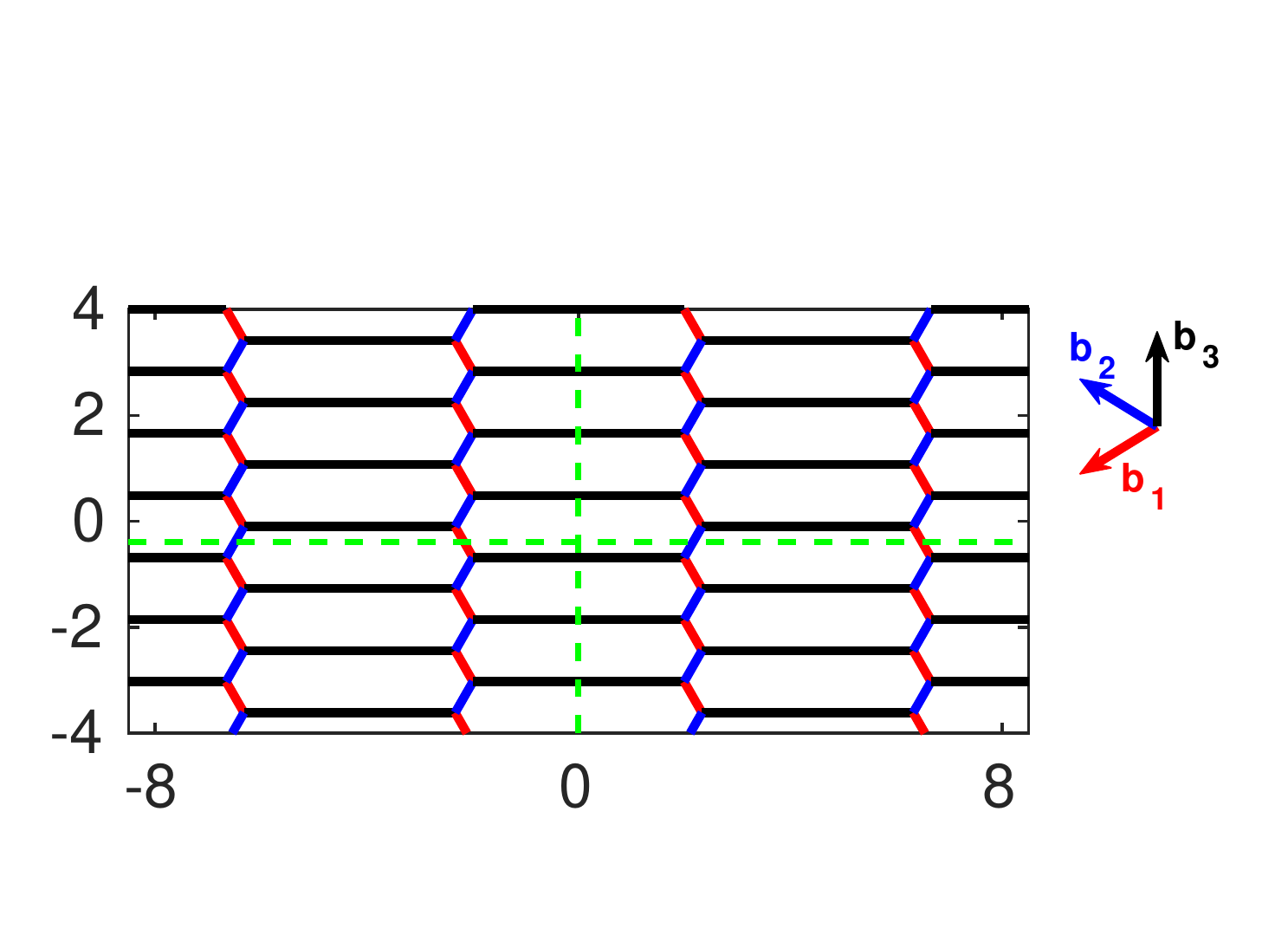}}
 \subfigure[]{    \includegraphics[width=0.40\linewidth]{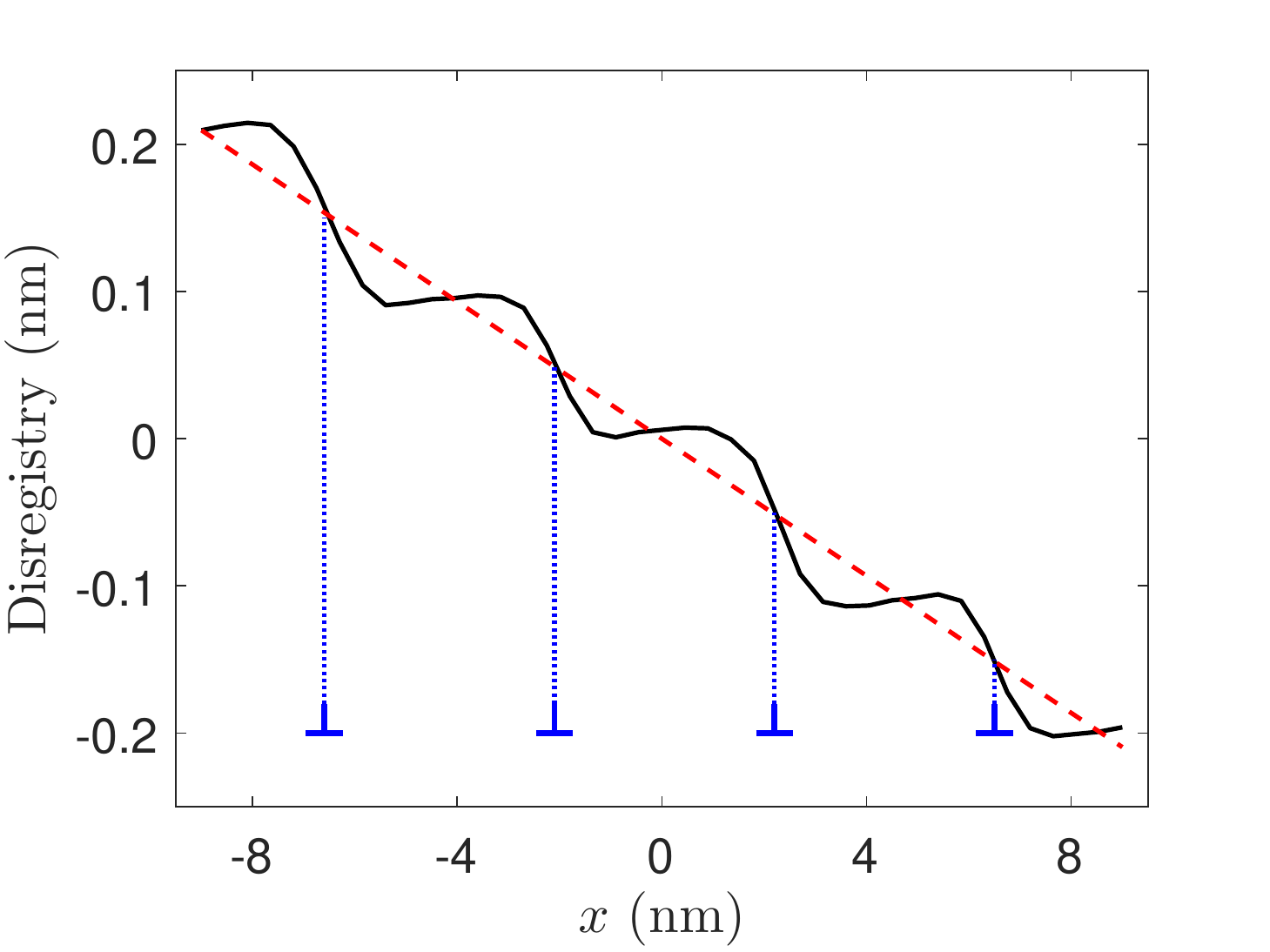}}
\subfigure[]{    \includegraphics[width=0.40\linewidth]{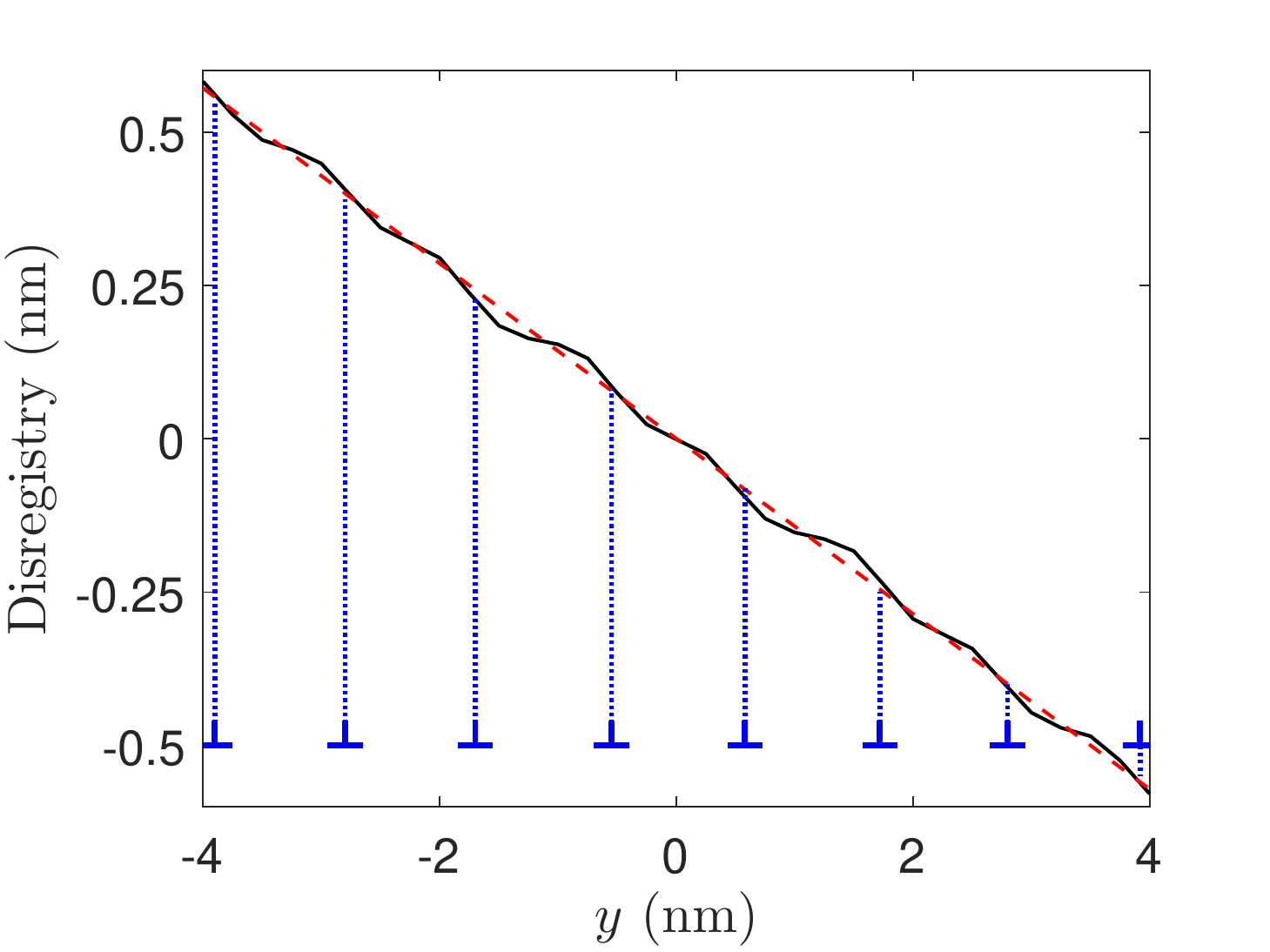}}
%\subfigure[]{     \includegraphics[width=0.35\linewidth]{hex.eps}}
     \caption{ (a)  Dislocation structure  calculated using our continuum model. The dislocation structure consists of a hexagonal network of three arrays of dislocations with Burgers vectors  $\mathbf{b}_1$, $\mathbf{b}_2$ and $\mathbf{b}_3$, shown by red, blue and black line segments, respectively. Unit: nm.
        (b) and (c) Disregistries along some cross-section lines obtained by atomistic simulations in Ref.~\cite{Wang201440} (reproduced from their data):
     (b) disregistry in the $x$ direction ($[11\bar{2}]_{\rm Cu}\|[1\bar{1}0]_{\rm Nb}$)  along a cross-section line parallel to the $x$ axis, and
     (c) disregistry in the $y$ direction ($[1\bar{1}0]_{\rm Cu}\|[001]_{\rm Nb}$)  along a cross-section line parallel to the $y$ axis.
          The black and red lines show the actual disregistry  and the unrelaxed uniform disregistry, respectively. The $\perp$ symbols show the locations of dislocations.
          %(e) Periodic parallelogram cell being employed in the derivation of method to identify the hexagonal network from dislocation densities and orientations obtained by our continuum model (see  Sec.~\ref{sec:model}).  Vectors $\mathbf a_1$ and $\mathbf a_2$ are the two sides and $A$ is the area of the periodic parallelogram unit cell.
           }
    \label{fig:Hex_CDP}
\end{figure}

In the dislocation network of this semicoherent interface obtained using our continuum model, dislocations with all the three Burgers vectors are present; see Fig.~\ref{fig:Hex_CDP}(a).
The converged values obtained using our continuum model are
%Simulation result using our continuum model shows that the dislocation structure of such an Cu$(111)$/Nb$(110)$ interface consists of a network of three arrays of dislocations with Burgers vectors  $\mathbf{b}_1$, $\mathbf{b}_2$ and $\mathbf{b}_3$ represented by the reciprocal vectors
$\mathbf N_1=(-0.1149,-0.0660)$/nm, $\mathbf N_2=(-0.1149,0.0660)$/nm and $\mathbf N_3=(0,0.8022)$/nm, respectively. This result  means that the dislocations with the three Burgers vectors have line
 directions of $\pmb \xi_1=(-0.4981,0.8671)$, $\pmb \xi_2=(0.4981,0.8671)$ and $\pmb \xi_3=(1,0)$ in the $xy$ plane, respectively,  and their densities are  $\rho_1=N_1=0.1325$/nm. $\rho_2=N_2=0.1325$/nm and $\rho_3=N_3=0.8022$/nm, respectively.
Recall that the local dislocation line direction is calculated by $\bm{\xi}_j = (\mathbf N_j / N_j)  \times \mathbf{n}$, where $\mathbf{n}$ is the unit normal vector of the interface. The $\mathbf b_3$-dislocations are edge dislocations, and the $\mathbf b_1$- and $\mathbf b_2$-dislocations are very close to edge, with an angle of $93^\circ$ between the dislocations and their Burgers vectors.

These three arrays of dislocations form a hexagonal network, because we have $\mathbf b_3=(-\mathbf b_1)+\mathbf b_2$ with $b_3^2<b_1^2+b_2^2$ and dislocation reaction from a  $\mathbf b_1$-dislocation and a $\mathbf b_2$-dislocation into a $\mathbf b_3$-dislocation is energetically favorable.  In this case, dislocations are disconnected segments instead of connected straight lines. Using the identification method presented at the end of  Sec.~\ref{sec:model} (Eq.~\eqref{eqn:segmentlength}), we can calculate the lengths of the three types of dislocation segments, which are $0.6642$nm, $0.6642$nm and $4.0208$nm
for the $\mathbf b_1$-, $\mathbf b_2$-, and $\mathbf b_3$-dislocation segments,  respectively. Based on these orientations and lengths of these dislocation segments, we can recover the hexagonal network as shown in Fig.~\ref{fig:Hex_CDP}(a).

%Considering the dislocation reaction in the network of three dislocations, the real interface dislocation network is a hexagonal network. We draw this network according to the derived reciprocal vectors $\mathbf{N}_j,j=1,2,3$ of three dislocation arrays in Fig.~\ref{fig:Hex_CDP}(a).
%The inter-dislocation distances obtained in our continuum model are $D_1=D_2 = 4.16 \rm {nm}$ and $D_3= 1.14\rm {nm}$.

We compare the dislocation structure obtained by our continuum model with the atomistic simulation result in Ref.~\cite{Wang201440}.  Disregistries along some cross-section lines obtained by atomistic simulations in Ref.~\cite{Wang201440} are shown in Figs.~\ref{fig:Hex_CDP}(b) and (c).  Figure~\ref{fig:Hex_CDP}(b) shows the disregistry in the $x$ direction ($[11\bar{2}]_{\rm Cu}\|[1\bar{1}0]_{\rm Nb}$)  along a cross-section line parallel to the $x$ axis, and this  cross-section line intersects with dislocations periodically with average distance of $4.40$nm between two neighboring intersecting points.
Figure~\ref{fig:Hex_CDP}(c) shows the disregistry in the $y$ direction ($[1\bar{1}0]_{\rm Cu}\|[001]_{\rm Nb}$)  along a cross-section line parallel to the $y$ axis, and this  cross-section line intersects with dislocations  periodically with average distance between two neighboring intersecting points  $1.14$nm.
On the other hand, in
 the dislocation structure obtained by our continuum model shown in Fig.~\ref{fig:Hex_CDP}(a),  the horizontal line  intersects with four $\mathbf b_1$- and $\mathbf b_2$-dislocations for $-8{\rm nm}\leq x\leq 8{\rm nm}$ and the distance between two neighboring intersecting points is $4.35$nm, and the vertical line
intersects with the array of the $\mathbf b_3$-dislocation segments with inter-dislocation distance of $1.15$nm.
These results agree excellently with those of atomistic simulations shown in Figs.~\ref{fig:Hex_CDP}(b) and (c) (see also Figs.~3-5 in Ref.~\cite{Wang201440}).

Dislocation structure on the NW semicoherent interface as well as the reference state have also be obtained in Ref.~\cite{Vattre2018} by using their continuum framework, in which a full anisotropic elasticity problem in the bulk of bicrystal is solved and the hexagonal
network on the interface with new dislocation segments with the third Burgers vector is formed by dislocation
reaction from the lozenge dislocation network of two sets of dislocations. The hexagonal networks obtained by using our continuum model and that by using their model (Fig.~6(d) in Ref.~\cite{Vattre2018})  are quite similar to each other. Quantitatively, across the same horizontal line as that in Fig.~\ref{fig:Hex_CDP}(a),
the distance between two neighboring intersecting points is $4.05$nm in their result (measured from their Fig.~6(d)), and  the distance between neighboring $\mathbf b_3$-dislocation segments is $1.13$nm (calculated from their data of the  lozenge dislocation network). These agreements between the results of the two models provide mutual validations of these models.

\subsection{Cu$(111)$/Nb$(110)$ interface with KS orientation relationship}

We also consider the Cu$(111)$/Nb$(110)$ interface with hexagonal pattern in the KS orientation relationship as shown in Fig.~\ref{fig:CDP_demo}(d), in which there is a relative rotation between the natural Cu fcc and Nb bcc lattices. The rotation angle of the natural Cu fcc lattice is $\theta_{\alpha} = 60^{\circ}$, and that of the natural Nb bcc lattice is $\theta_{\beta}=54.74^{\circ}$. The rotation angle of the reference lattice  is approximately chosen as the average of $\theta_{\alpha}$ and $\theta_{\beta}$, i.e., $\theta=(\theta_{\alpha}+\theta_{\beta})/2$.
The three possible Burgers vectors and the distortion matrices of this interface can be calculated by rotations from the corresponding ones in the interface with the NW orientation relationship, using Eqs.~\eqref{eqn:BV_rotate}, \eqref{eqn:S_RCDP} and \eqref{eqn:BV_hex}, \eqref{eqn:S_hex_CDP}.

\begin{figure}[htbp]
 \centering
%    \includegraphics[width=0.45\linewidth]{Hex_RCDP_MD.jpg}
%\subfigure[]{  \includegraphics[width=0.32\linewidth]{Hex_RCDP_demo.eps}}
   \includegraphics[width=0.55\linewidth]{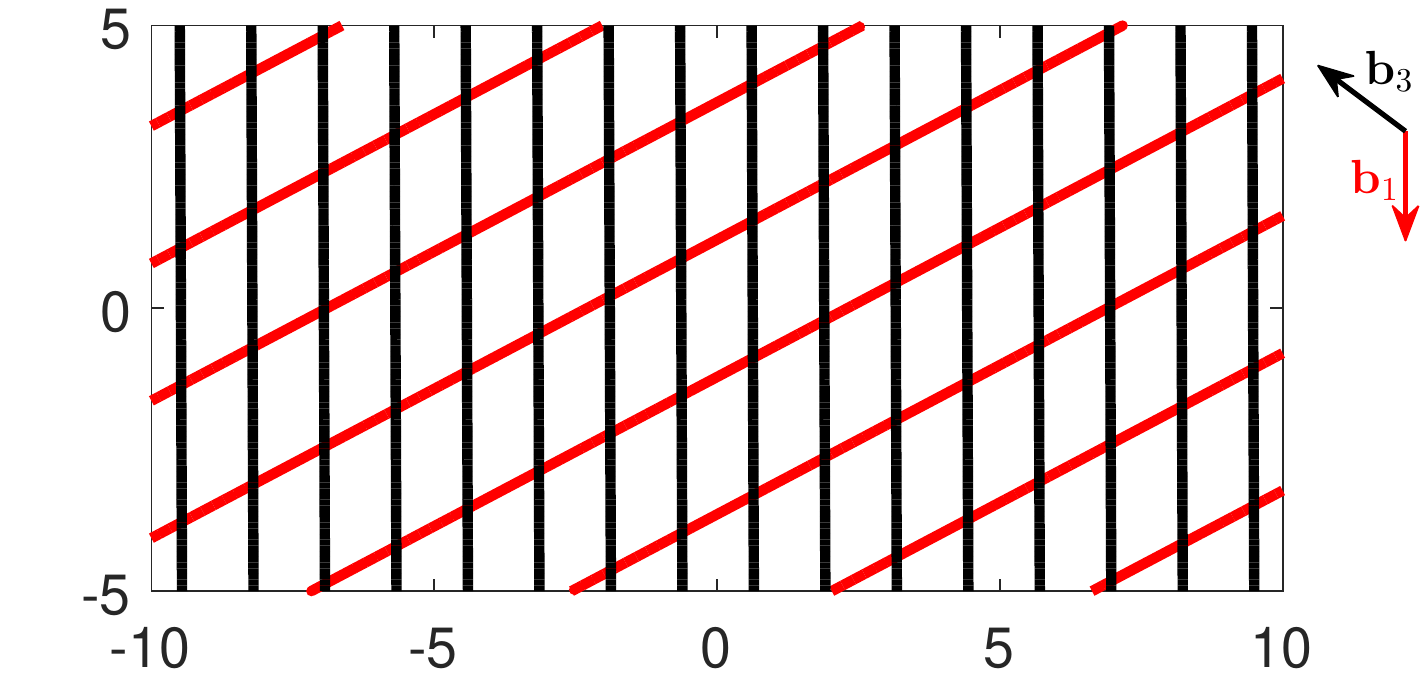}
    \caption{Dislocation structure of the Cu$(111)$/Nb$(110)$ interface in KS OR calculated using our continuum model. The dislocation structure consists of a parallelogram network of dislocation arrays with Burgers vectors $\mathbf{b}_1$ and $\mathbf{b}_3$, shown by red and black lines, respectively.}
    \label{fig:Hex_RCDP}
\end{figure}

Simulation result using our continuum model shows that the dislocation structure of this Cu$(111)$/Nb$(110)$ interface with  KS orientation relationship consists of a parallelogram network of two arrays of dislocations with Burgers vectors  $\mathbf{b}_1$ and $\mathbf{b}_3$ represented by the reciprocal vectors $\mathbf N_1=(0.2173,-0.4118)/$nm and $\mathbf N_3=(-0.7913,-0.0032)/$nm, respectively; see Fig.~\ref{fig:Hex_RCDP} for the obtained dislocation network.
These results of the continuum model mean that the two arrays of dislocations have line directions $\pmb \xi_1=(-0.8844,-0.4667)$ and $\pmb \xi_2=(-0.0040, 1.0000)$ with inter-dislocation distances  $D_1 = 2.148 \rm {nm}$ and $D_3= 1.264\rm {nm}$, respectively. Accordingly, the angles between the two arrays of dislocations and the $x$ axis are  $-152.2^\circ$ and $90.2^\circ$, respectively.
 We compare this dislocation structure with that obtained by atomistic simulation in Ref.~\cite{Wang201440} (Figs.~8 and 9 in Ref.~\cite{Wang201440}), and observe excellent agreement between the two results.
In the  atomistic simulation result in Ref.~\cite{Wang201440}, the inter-dislocation distances in the two arrays of dislocations are $D_1=2.131 \rm {nm}$, $D_3= 1.245 \rm {nm}$ and the dislocation line directions have angles $152^{\circ}$ and $90^{\circ}$ with respect to the $x$-axis, respectively (see Table 2 in Ref.~\cite{Wang201440}. Notice that the $+z$ direction is pointing downward there).

\section{Conclusions}\label{sec:conclusion}
 In summary, we have developed a continuum model for the dislocation structures of semicoherent interfaces based on constrained energy minimization.
 In our model,
the dislocation structure of a semicoherent interface is obtained by minimizing the
energy of the equilibrium dislocation network with respect to  all the possible Burgers vectors, subject to the constraint of the Frank-Bilby equation.
Comparisons with atomistic simulation results and results of other available models show that
our continuum model is able to give excellent predictions of dislocation structures on semicoherent interfaces.
%The continuum model is validated by comparisons with atomistic simulation results.

\section*{Acknowledgements}
 This work was supported
by the Hong Kong Research Grants Council General Research Fund through grant 16302818.

%The classical Frank-Bilby equation that governs the dislocation structures on
%semicoherent interfaces is not able to determine a unique solution. The available
%methods in the literature either use further information from atomistic simulations or
%consider only special cases where the structure is uniquely given by the Frank-Bilby
%equation (dislocations with no more than two Burgers vectors).

\section*{Data availability}
The datasets generated in study are available upon %reasonable
 request.

\section*{References}
\bibliographystyle{elsarticle-num}
\bibliography{proposal}

\end{document}